\pdfoutput=1
\documentclass[twocolumn]{aastex63}
\usepackage{multirow}
\usepackage{CJKutf8}

\usepackage{subfigure}
\usepackage{amsmath}
\usepackage{graphicx}
\usepackage{changepage} 
\usepackage{hyperref}
\usepackage{hhline}
\newcommand{\galex}{\emph{GALEX}}
\newcommand{\energyratio}{$\mathcal{R}_E$}
\newcommand{\peakratio}{$\mathcal{R}_{\lambda L \lambda}$}
\newcommand{\msun}{\hbox{M$_{\odot}$}}
\newcommand{\teff}{\hbox{$T_{\rm eff}$}}

\begin{document}
\begin{CJK*}{UTF8}{gbsn}
\title{Stellar Flares Are Far-Ultraviolet Luminous}
\shorttitle{Stellar Flares Are Far-Ultraviolet Luminous}

\submitjournal{MNRAS}

\shortauthors{Berger et al.}

\correspondingauthor{Vera Berger}
\email{vlb36@cam.ac.uk}

\author[0000-0002-7303-8144]{Vera L. Berger}
\altaffiliation{Churchill Scholar}
\affiliation{Cavendish Laboratory, Department of Physics, University of Cambridge, JJ Thomson Avenue, Cambridge, CB3 0HE, UK}
\affiliation{Institute for Astronomy, University of Hawai`i, 2680 Woodlawn Drive, Honolulu, HI 96822, USA}
\affiliation{Department of Physics and Astronomy, Pomona College, 333 N. College Way, Claremont, CA 91711, USA}

\author[0000-0001-9668-2920]{Jason T. Hinkle}
\altaffiliation{FINESST FI}
\affiliation{Institute for Astronomy, University of Hawai`i, 2680 Woodlawn Drive, Honolulu, HI 96822, USA}

\author[0000-0002-2471-8442]{Michael A. Tucker}
\altaffiliation{CCAPP Fellow}
\affiliation{Department of Astronomy, The Ohio State University, 140 West 18th Avenue, Columbus, OH 43210, USA}
\affiliation{Center for Cosmology and Astroparticle Physics, The Ohio State University, 191 W.~Woodruff Avenue, Columbus, OH 43210, USA}

\author[0000-0003-4631-1149]{Benjamin J. Shappee}
\affiliation{Institute for Astronomy, University of Hawai`i, 2680 Woodlawn Drive, Honolulu, HI 96822, USA}

\author[0000-0002-4284-8638]{Jennifer L. van Saders}
\affiliation{Institute for Astronomy, University of Hawai`i, 2680 Woodlawn Drive, Honolulu, HI 96822, USA}

\author[0000-0001-8832-4488]{Daniel Huber}
\affiliation{Institute for Astronomy, University of Hawai`i, 2680 Woodlawn Drive, Honolulu, HI 96822, USA}
\affiliation{Sydney Institute for Astronomy (SIfA), School of Physics, University of Sydney, NSW 2006, Australia}

\author[0000-0003-4739-1152]{Jeffrey W. Reep}
\affiliation{Institute for Astronomy, University of Hawai`i, 34 Ohia Ku St., Pukalani, HI 96768, USA}

\author[0000-0003-4043-616X]{Xudong Sun (孙旭东)}
\affiliation{Institute for Astronomy, University of Hawai`i, 34 Ohia Ku St., Pukalani, HI 96768, USA}

\author[0000-0002-7663-7652]{Kai E. Yang (杨凯)}
\affiliation{Institute for Astronomy, University of Hawai`i, 34 Ohia Ku St., Pukalani, HI 96768, USA}

% --------------- ABSTRACT ---------------------------------------
\begin{abstract}
We identify 182 flares on 158 stars within 100~pc of the Sun in both the near-ultraviolet (NUV: $1750-2750$ \AA) and far-ultraviolet (FUV: $1350-1750$ \AA) using high-cadence light curves from the \textit{Galaxy Evolution Explorer (GALEX)}.
Ultraviolet (UV) emission from stellar flares plays a crucial role in determining the habitability of exoplanetary systems. However, whether such UV emission promotes or threatens such life depends strongly on the energetics of these flares.
Most studies assessing the effect of flares on planetary habitability assume a $9\,000$~K blackbody spectral energy distribution that produces more NUV flux than FUV flux ($\mathcal{R} \equiv F_{\rm FUV} / F_{\rm NUV} \approx \frac{1}{6}$). Instead, we observe the opposite with the excess FUV reaching $\mathcal{R} \approx \frac{1}{2} \-- 2$, roughly $3\--12$ times the expectation of a $9\,000$~K blackbody.
The ratio of FUV to NUV time-integrated flare energies is 3.0 times higher on average than would be predicted by a constant $9\,000$~K blackbody during the flare.  
Finally, we find that the FUV/NUV ratio at peak tentatively correlates ($\sim2 \sigma$ significance) both with total UV flare energy and with the $G - RP$ color of the host star.
On average, we observe higher FUV/NUV ratios at peak in $E_{\text{UV}}>10^{32}$~erg flares and in flares on fully convective stars.
\end{abstract}


% --------------- INTRO ---------------------------------------
\section{Introduction}\label{s1:intro}
\end{CJK*}
Stellar flares are energetic bursts of electromagnetic radiation driven by magnetic reconnection \citep[e.g.,][]{litvinenko99, benz10, shibata16}. 
Flares are more frequently observed on low-mass stars ($M<1.5\msun$), which have surface convection zones \citep[e.g.,][]{petterson89, balona15, davenport16, doorsselaere17}. Flares on M-dwarf stars are of particular interest because of these stars' heightened magnetic activity and ideal candidacy to detect exoplanets in the habitable zone \citep[e.g.,][]{endl03, kaltenegger09, reiners12}.

Ultraviolet emission from flares impacts the habitability of exoplanets orbiting flaring stars. 
\citet{rimmer18} delineated ``abiogenesis zones" in which flare rates and energies could deliver sufficient UV photons to drive prebiotic chemistry. 
On the other hand, \citet{tilley19} characterized  ``ozone depletion zones" in which M-dwarf flares of sufficient rates and energies deplete the ozone column of habitable exoplanets.
Flares with energies above $10^{34}$~erg may contribute to either effect. Thus, for 
sufficiently high flare rates, it is unclear exactly where flares help versus hinder the development of complex molecules. 

\begin{figure*}
  \centering
  \subfigure[]{\includegraphics[width=0.485\textwidth]{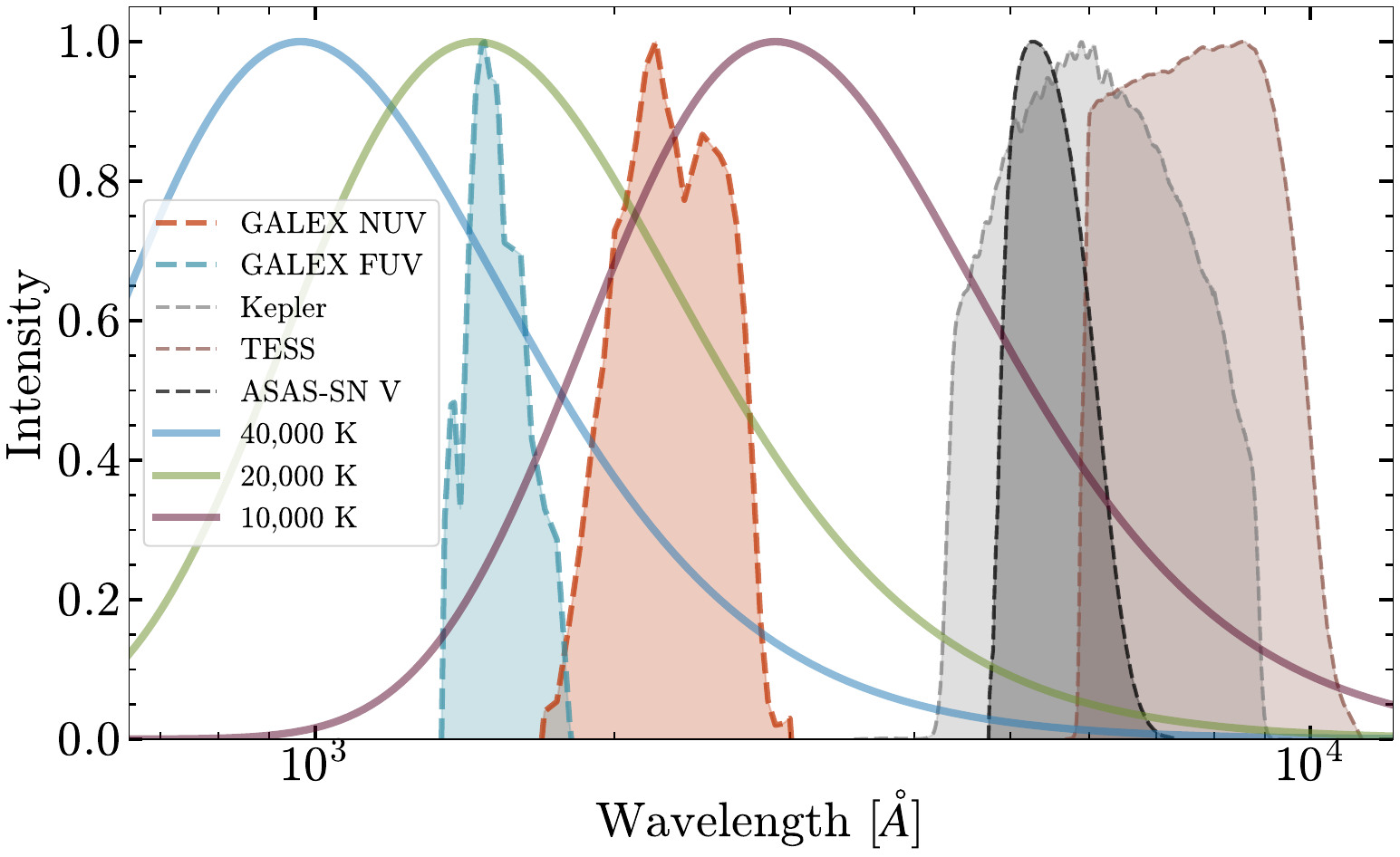}\label{fig:filter_resp}}
  \hfill
  \subfigure[]{\includegraphics[width=0.49\textwidth]{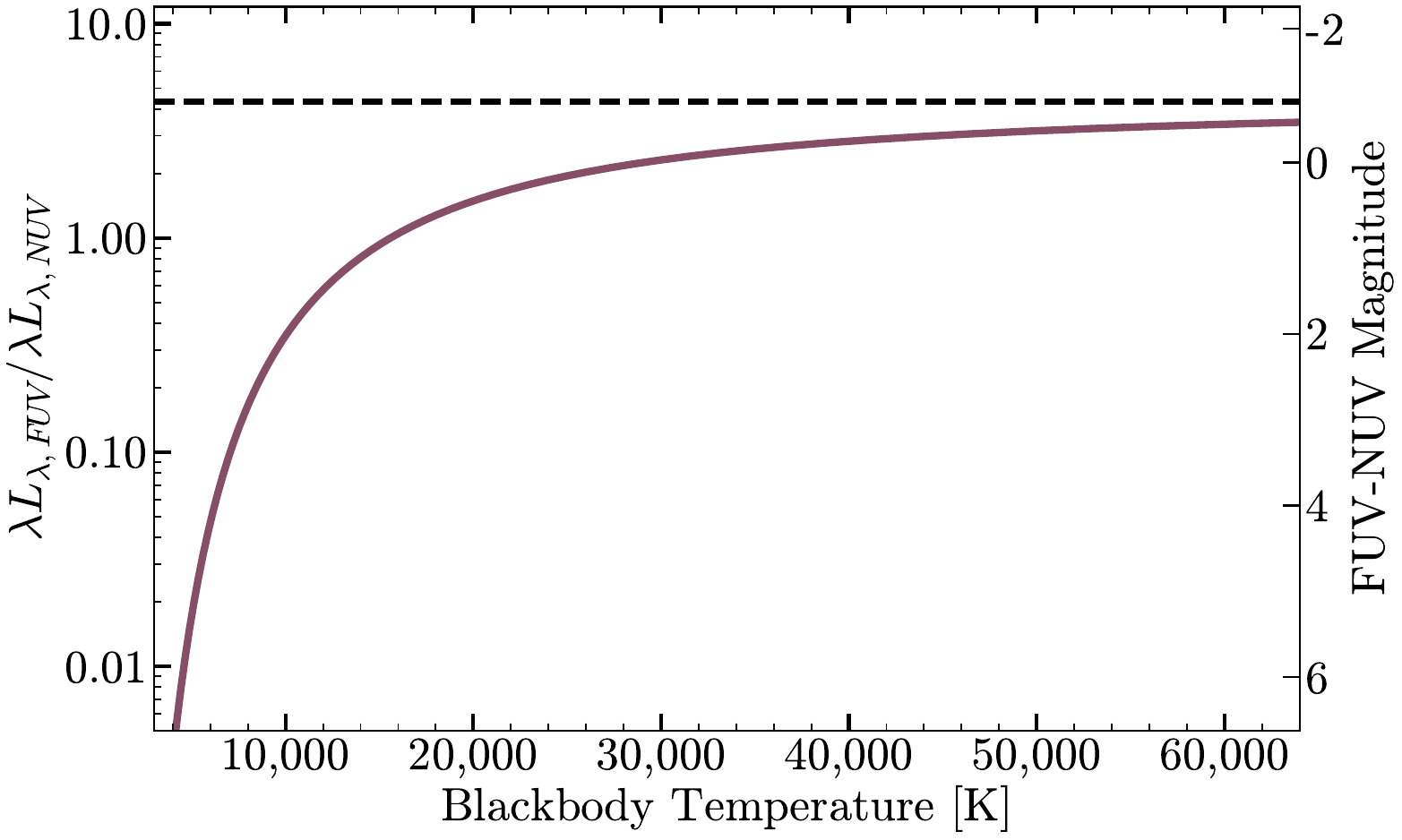}\label{fig:color-temp}}
  
  \caption{\emph{Left panel:} Normalized blackbody curves (solid lines) spanning a range of observed temperatures plotted against filter response functions (dashed lines) for telescopes commonly used for flare studies. 
  The TESS (brown), \emph{Kepler} (black), and ASAS-SN V band (gray) response functions fall on the Rayleigh-Jeans tail for temperatures above $\sim 7\,000$~K. \galex \, NUV (red) and FUV (teal) filter response functions fall around the peaks of blackbody SEDs for temperatures $\sim 10\,000 \-- 30\,000$~K. \emph{Right panel:} Relation between blackbody temperature and FUV/NUV $\lambda L_{\lambda}$ ratio, computed using synthetic photometry. The dashed line at 4.33 denotes the numerically-computed asymptote once the \galex{} filters are both in the Rayleigh-Jeans limit.}
  \label{fig:fig1}
\end{figure*}

Recent optical studies have found few stars that display superflare ($E > 10^{33}$~erg) rates and energies sufficient to affect exoplanet habitability \citep{schmidt14, schmidt16, schmidt19, rodriguez18, rodriguez20, gunther20, feinstein20, zeldes21, bogner21}. These studies typically assume a constant-temperature blackbody spectrum for flares ($\teff \approx 9\,000 - 10\,000$~K) to estimate the incident UV flux. Given that a $9\,000$~K blackbody produces 84\% less FUV emission than NUV emission, small temperature deviations from the adopted spectral energy distribution (SED) can produce large variations in the incident UV flux. 

While large-scale observational studies of stellar flares have primarily been conducted in the optical wavelengths \citep[e.g.,][]{walkowicz11, shibayama13, hawley14, davenport16,  schmidt19, rodriguez20, gunther20, feinstein20}, these wavelengths can comprise a small fraction of a flare's total emission \citep{maehara12, namekata17}.  
As displayed in Figure \ref{fig:filter_resp}, the filter responses for the \emph{Kepler} space telescope \citep{kepler}, the Transiting Exoplanet Survey Satellite \citep[\emph{TESS,}][]{tess} and the All-Sky Automated Survey for Supernovae \citep[ASAS-SN,][]{asassn_shappee, asassn_kochanek, asassn_hart} capture only the Rayleigh-Jeans tail of blackbody emission for $\teff > 7\,000$~K.
In addition, NUV and optical flare emission may arise from M-dwarf stars' chromospheres and upper photospheres \citep{joshi21}, whereas
far-ultraviolet emission stems from the upper chromosphere and may correspond to the flare's impulsive phase of heating and compression of plasma \citep[]{neupert68, dennis93, hawley03, benz10}. 

While NUV and optical emission may be well-represented by a blackbody, some superflares have exhibited significant line emission in the FUV,
sometimes surpassing quiescent flux levels by a factor of 100 \citep{france16}. 
\citet{loyd18a} computed a blackbody temperature of $15\,500$~K for the ``Hazflare" in FUV spectra from the HST Cosmic Origins Spectrograph (COS, $1170 - 1430$~\AA). This flare exhibited substantial enhancements in C III, Si IV, S III, and N V emission lines; however, as with the 1985 Great Flare on AD Leo \citep{hawley91}, a hot blackbody continuum dominated the total FUV emission.
Furthermore, \citet{froning19} obtained a far/extreme-UV spectrum of the M dwarf GJ 674 with HST COS ($1065-1365$~\AA) during a flare, revealing a blackbody-dominated spectrum with $\teff \approx 40\,000$~K and numerous superimposed C, Si, N, and Fe emission lines.

\begin{figure*}
  \centering
  \subfigure[]{\includegraphics[width=0.46\textwidth]{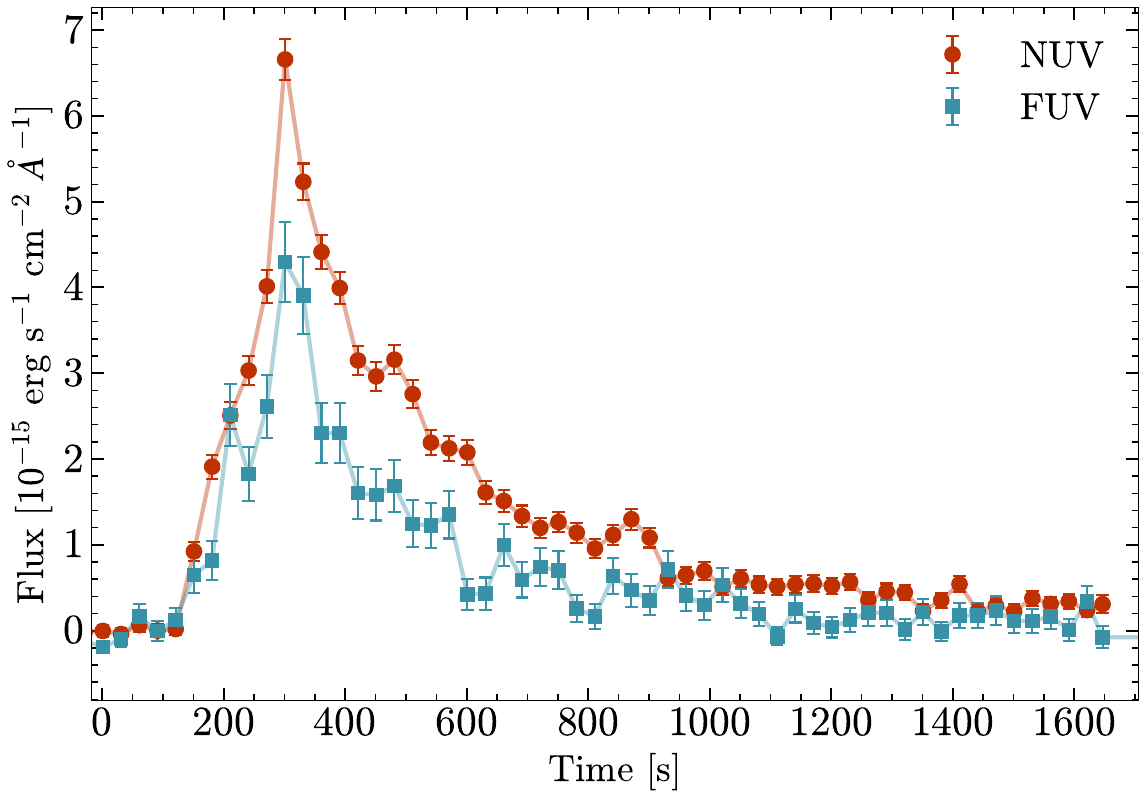}\label{fig:flarelc_lowerratio}}
  \hfill
  \subfigure[]{\includegraphics[width=0.464\textwidth]{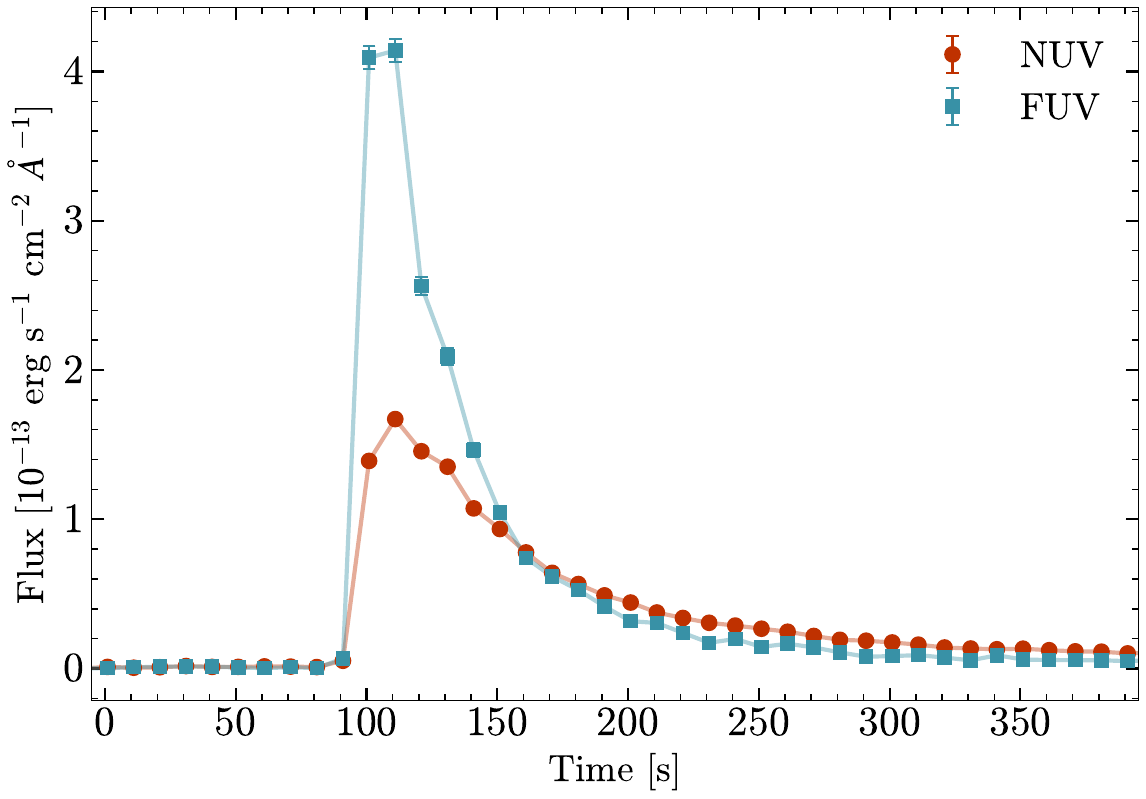}\label{fig:flarelc_highratio}}
  
  \vspace{1em}
  
  \subfigure[]{\includegraphics[width=0.46\textwidth]{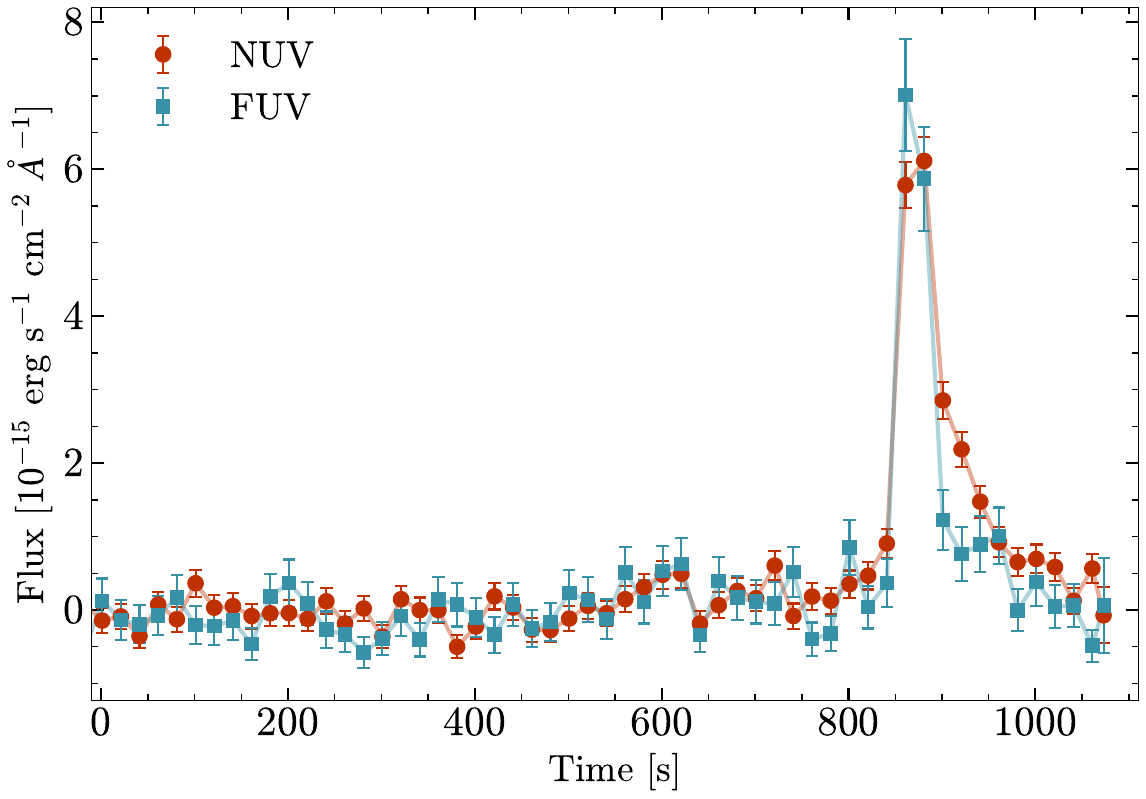}\label{fig:randflare1}}
  \hfill
  \subfigure[]{\includegraphics[width=0.468\textwidth]{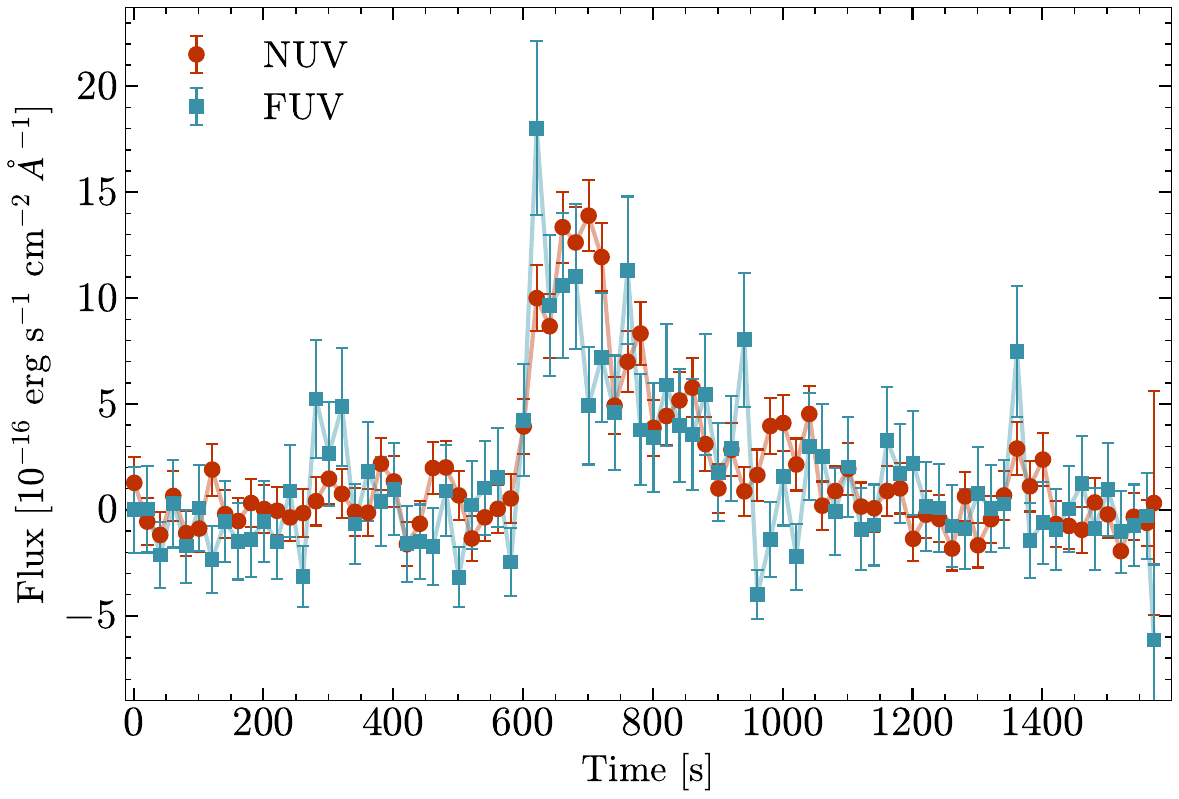}\label{fig:randflare2}}
  
  \caption{Example NUV and FUV light curves for two flares in our sample that are representative of the range of FUV/NUV flux and SNR we observe. \emph{Upper panel:} Two high SNR flares spanning the range of FUV/NUV fluxes present in our sample. On the left is Gaia 3537511712697046656, a K6V star 56 pc from Earth. On the right is Gaia 1461125613285603840, an M4V star 13 pc from Earth. \emph{Lower panel:} Flares chosen randomly from our sample. On the left is Gaia 365548347350541952, an M1.5V star 22 pc from Earth. On the right is Gaia 3858896247076028288, an M3.5V star 15 pc from Earth.}
  \label{fig:flarelcs}
\end{figure*}

In contrast, FUV spectra of the Sun show flare emission dominated by line emission instead of a thermal continuum \citep{brekke96, simoes19}. 
Given the intrinsic differences between M-dwarf and solar magnetic fields  \citep[and thus flare activity;][]{gunther20}, it remains unclear when line emission dominates or continuum dominates for a given flare.

A more direct method for studying the impact of flares on habitability is time-resolved UV photometry of stars during flares. The NASA \textit{Galaxy Evolution Explorer (GALEX)}  space mission \citep{galex05, galex07_morrissey} provides a unique opportunity to study ultraviolet emission from flares.
\galex \, simultaneously observed in the far-ultraviolet (FUV: $1350-1750$ \AA) and near-ultraviolet (NUV: $1750-2750$ \AA) bands \citep{galex05}. 
In addition, the micro-channel plate detectors aboard \emph{GALEX} recorded the timing of individual photon events to $\approx 5~\rm{ms}$ precision. 
We can therefore construct high-cadence light curves with gPhoton \citep{million16} to process \textit{GALEX} observations at arbitrary time resolution.

Early flare studies utilizing the time-resolved UV photometry of \textit{GALEX}  have been limited to small sample sizes. 
Studies of a few flares using simultaneous \galex{} FUV and NUV observations 
have characterized increases in FUV/NUV flux during superflares reaching a maximum ratio of 13 in the most extreme case, a factor of $>50$ higher than would be predicted by a $9\,000$~K blackbody \citep{robinson05, welsh06}. 
FUV emission from host stars can photolyze CO$_2$ in planetary atmospheres, producing atomic oxygen which can then recombine to form O$_2$ \citep{tian14}. Therefore, high ratios of FUV/NUV radiation such as that observed by these studies may produce sufficient abiotic atmospheric oxygen to constitute a false biosignature \citep{tian14, harman15}. In addition, high FUV/NUV ratios suggest that that the typical SED assumptions for flare studies may be underestimating the true energy and ionizing flux of superflares.

\begin{table*}[]
    \centering
    \begin{tabular}{p{3.2cm} p{1.5cm} p{1.5cm} p{2cm} p{1cm}  p{1cm} p{1.5cm}}
    \hhline{=======}
\multirow{2}{7em}{Gaia EDR3 ID} & \multirow{2}{2em}{RA [deg]}  & \multirow{2}{2em}{Dec [deg]}  & \multirow{2}{3em}{Distance [pc]}   & \multirow{2}{2em}{$M_\text{G}$ [mag]}  & \multirow{2}{2em}{$RP$ [mag]} &  Spectral Type\\ 
\hline

1013481650428107648 & 132.6026 & 46.3209 & $81.95 \pm 0.15$ & 14.22 & 13.09  & M3V \\ 
1027636694403121664 & 130.6758 & 51.4823 & $60.10 \pm 0.07$ & 13.92 & 12.73 & M3V \\ 
1047687216648750976 & 159.5389 & 59.4848 & $93.44 \pm 0.56$ & 16.77 & 15.41  &  M5V \\ 
1058559634300870400 & 162.0490 & 64.7243 & $96.62 \pm 0.21$ & 15.21 & 13.96   &  M4V \\ 
1059924128230565888 & 161.0179 & 67.0126 & $47.40 \pm 0.03$ & 13.66 & 12.50   &  M3V \\ 
1070172916631499520 & 150.0283 & 68.1115 & $76.99 \pm 0.14$ & 15.44 & 14.22  &  M3.5V\\ 
1073027695493625856 & 157.6594 & 68.5490 & $88.37 \pm 0.14$ & 14.39 & 13.19    &  M3.5V \\ 
1076599905693041280 & 156.8567 & 71.0655 & $95.87 \pm 0.26$ & 15.39 & 14.17   &  M3.5V \\ 
1113525010848246784 & 96.6996 & 71.1834 & $90.77 \pm 0.42$ & 16.80 & 15.50    &  M4.5V \\
1119235182751380736	& 140.6038 & 70.1558	& $91.59 \pm 1.32$ & 18.56 & 17.11 & M6V \\
$\cdots$ & $\cdots$ & $\cdots$ & $\cdots$ &$\cdots$ & $\cdots$ & $\cdots$  \\
\hline 
    \end{tabular}
    \caption{Star properties; full table available in the online version of the manuscript.}
    \label{tab:stars}
\end{table*}

In this work, we investigate all stars within 100 pc that were observed by \textit{GALEX} simultaneously in the NUV and FUV. 
We use a sample of 182 flares on 158 stars and show that this FUV excess emission is not restricted to superflares but a general characteristic of flares on low-mass stars.

In Section \ref{s2:sample} we provide an overview of our sample and flare detection method. 
Section \ref{s3:fuv} characterizes FUV and NUV flare emission in comparison to blackbody temperatures. Section \ref{s4:discussion} discusses the implications of our findings for characterizing the impact of stellar flares on the habitability of orbiting exoplanets.

% % --------------- Sample? ---------------------------------------
\vspace{1em}
 \section{Sample Selection and Flare Detection}\label{s2:sample}

We select our targets from the Gaia Catalogue of Nearby Stars (GCNS), which is complete down to spectral type M8 within 100 pc \citep{gaiaedr3}.
We exclude sources within one FWHM ($\approx 5.6$ arcsec) of another star in the GCNS to avoid contamination. 
We then use the gPhoton database and processing software \citep{million16} to search for \galex{} coverage of stars in the GCNS and perform aperture photometry. 
We exclude data with flags from gPhoton corresponding to aperture or annulus events in pixels that are contiguous to a masked hotspot or the detector edge.
We bin photon events at cadences scaled by the physical distance of the corresponding sources. 
Shorter bins are employed for nearby sources. For white-light flares, amplitude and duration are correlated \citep{maehara15}, and so we expect shorter flares to be fainter and thus only be visible in more nearby stars. More distant objects are binned with longer cadences to enhance signal-to-noise ratios.
We use 10, 20, and 30 s bins for stars within 15 pc, between 15 - 50 pc, and between $50 - 100$ pc respectively.

We find that bright sources exhibit significant systematics. We remove any light curves with median flux brighter than 15.4 mag NUV, a limit established by examining 2000 randomly chosen light curves by eye. Visual examination of stellar images and their background light curves suggests that contaminated sources either have a high cross-correlation between background and source flux over time, or a high standard deviation in background flux.  We minimize this contamination by also removing light curves with background flux with a median above 40 counts or standard deviation above 6 counts.

\begin{deluxetable*}{ccccccc}[htbp!]

\tablecaption{Flare properties; full table available in the online version of the manuscript. \label{tab:flares}}
\tablehead{
    \colhead{Gaia EDR3 ID} &
    \colhead{t$_\text{start}$} &
    \colhead{t$_\text{end}$} &
    \colhead{Exp. Time} &
    \colhead{\energyratio} &
    \colhead{\peakratio} &
    \colhead{UV Energy} \\
    \colhead{}  &
    \colhead{[\emph{GALEX} s]} &
    \colhead{[\emph{GALEX} s]} &
    \colhead{[s]} &
    \colhead{} &
    \colhead{} &
    \colhead{[erg]}
}
\startdata
1013481650428107648 & 794114866 & 794115011 & 145 & $0.57 \pm 0.06$ & $1.16 \pm 0.09$ & 1.96$^{+0.10}_{-0.10}$e+32  \\
1027636694403121664 & 823888055 & 823888475 & 420 & $0.39 \pm 0.03$ & $0.75 \pm 0.04$ & 2.93$^{+0.09}_{-0.09}$e+32  \\
1047687216648750976 & 890596076 & 890596136 & 60 & $0.84 \pm 0.21$ & $1.16 \pm 0.21$ & 4.70$^{+0.60}_{-0.60}$e+31  \\
1058559634300870400 & 758796977 & 758797577 & 600 & $0.78 \pm 0.04$ & $0.65 \pm 0.05$ & 4.00$^{+0.17}_{-0.17}$e+32  \\
1059924128230565888 & 820629087 & 820629182 & 100 & $0.81 \pm 0.23$ & $1.20 \pm 0.39$ & 9.03$^{+1.44}_{-1.37}$e+30  \\
1069666346712783488 & 882307565 & 882308342 & 777 & $0.50 \pm 0.07$ & $0.46 \pm 0.14$ & 2.28e$^{+0.16}_{-0.16}$+32  \\
1070172916631499520 & 820545790 & 820545880 & 90 & $0.30 \pm 0.16$ & $0.52 \pm 0.21$ & 2.09e$^{+0.35}_{-0.35}$+31  \\
1073027695493625856 & 882461389 & 882462136 & 748 & $0.48 \pm 0.02$ & $0.92 \pm 0.03$ & 3.06e$^{+0.04}_{-0.04}$+33  \\
1076599905693041280 & 859394503 & 859394653 & 150 & $0.48 \pm 0.08$ &$ 0.59 \pm 0.09$ & 1.31e$^{+0.10}_{-0.10}$+32  \\
1113525010848246784 & 912762216 & 912762336 & 120 & $0.74 \pm 0.15$ & $1.54 \pm 0.23$ & 7.87$^{+0.82}_{-0.84}$e+31 \\
$\cdots$ &$\cdots$ &$\cdots$ &$\cdots$ & $\cdots$ & $\cdots$ & $\cdots$  \\
\enddata
\end{deluxetable*}

We use light curves in the NUV for flare detection because the \emph{GALEX} NUV detector has better throughput and efficiency \citep{bianchi14}.
The flare detection approach relies on finding data points that are inconsistent with noise in the light curve and representative of true source variability \citep[e.g.,][]{chang15}.
 We estimate a baseline flux by iteratively sigma-clipping the light curve 2$\sigma$ above the median flux.
 Given a time period $[t_a,t_b]$ of a light curve, we compute the mean flux $\bar f_{ab}$ and standard deviation $ \sigma_{ab}$.
For the $i$th flux measurement with calculated uncertainty $\sigma_i$ within the time period, we evaluate:

\begin{equation}
    \frac{f_i - \bar f_{ab}}{\sqrt{\sigma_{i}^2 + \sigma_{ab}^2}} \geq n,
   \label{eq:condition1}
\end{equation}

\noindent where a flux satisfying Equation \ref{eq:condition1} corresponds to an $n\sigma$ detection.
We first search for an individual $3\sigma$ detection in the light curve and then look to the neighboring points in the range  $[t_{i-2}, t_{i+2}]$, excluding $t_i$. 
If there are two adjacent points in this range corresponding to $2\sigma$ detections we then flag the visit as containing a flare.
Finally, we require that the flux does not peak at the first or last photometric point in the visit since some objects show decreasing or increasing trends only at the very beginning or end of a visit.

For this study we only include flares with observations in both FUV and NUV to enable the calculation of color temperatures.  A full analysis of all light curves, including NUV-only flares and injection-recovery tests, is reserved for a forthcoming paper (Berger et al. in prep).
As a final cut, we require that the FUV light curve has fewer than 3 missing observations within 5 points of peak NUV flare brightness. 
This method produces 188 candidates. 
Even with our quality cuts, six additional sources display variability that we classify as a false-positive upon examining the light curves and 2D images. These candidates appear within crowded fields.

Our final sample is composed of 182 flares with simultaneous observations in the NUV and FUV.  We present their sources in Table \ref{tab:stars}.  Spectral classification of our sample is discussed in Section~\ref{3.3} but the sample is composed of 79\% fully convective stars ($\geq$~M3, $n=143$), 16\% partially convective M stars ($n=30$), and 5\% K stars ($n=9$). 
Figure \ref{fig:flarelcs} displays the range of light curves we observe; the top panel shows two high signal-to-noise flares that span the range of FUV/NUV flux we observe, and the lower panel shows two representative randomly selected flares in our sample.

% % --------------- S ---------------------------------------
\section{FUV Flare Emission}\label{s3:fuv}

In this section we calculate FUV/NUV emission ratios, inferred temperatures and UV energies for the 182 flares in our sample, summarized in Table \ref{tab:flares}, then compare the inferred blackbody color temperatures to the typically-assumed $9\,000$~K blackbody SED for flares.

\subsection{Flare temperatures assuming blackbody emission}\label{sec:color_temp}
Under the assumption that the spectral energy distribution of a flare can be well-represented as a blackbody, we can construct a one-to-one relation between blackbody temperature and FUV/NUV $\lambda L_{\lambda}$ ratio.  
We do this by integrating the transmission function and blackbody SED over the range of wavelengths covered by each filter \citep{koornneef86}.
We assume interstellar extinction is negligible since our sources are within the Local Bubble \citep[e.g.,][]{cox87, fossati17}, characterized by a low-density interstellar medium. 
Figure \ref{fig:color-temp} shows the FUV/NUV $\lambda L_{\lambda}$ ratio for a range of blackbody temperatures computed using synthetic photometry.
Note that the color-temperature curve flattens considerably around $50\,000$~K, where both \galex{} filters land on the Rayleigh-Jeans tail of the blackbody and thus the FUV-NUV color can no longer distinguish between hotter temperatures.
The color-temperature curve asymptotes to FUV/NUV $= 4.33$ for arbitrarily high \teff.
 
We estimate the inferred blackbody temperature for each observation during a flare by directly measuring the FUV/NUV $\lambda L_{\lambda}$ ratio. 
Uncertainties on the color temperature are estimated by propagating the uncertainties on observed fluxes.

\subsection{Excess FUV emission}
We estimate the FUV excess flux for each observation during a flare by scaling a $9\,000$~K blackbody to the observed NUV flux. Then, we impute the expected FUV emission for comparison to our observations. We denote the ratio of time-integrated energies by \energyratio$= E_{\rm FUV} / E_{\rm NUV}$.

A $9\,000$~K blackbody produces a time-integrated energy ratio of 0.17 whereas we find a median \energyratio{}  = 0.50 ($T_{\rm eff} \approx 13\,500$~K) and a maximum \energyratio{}  = 1.06 ($T_{\rm eff} \approx 21\,000$~K). 
Thus, a $9\,000$~K blackbody underestimates the median observed \energyratio{} by a factor of 3.0 and underestimates the maximum ratio by a factor of 6.3.
Among the flares in our sample, 98\% ($n=178$) display \energyratio{} ratios that exceed expectations for a $9\,000$~K blackbody.
This suggests that a constant $9\,000$~K blackbody SED is insufficient to account for the levels of FUV emission we observe.

\begin{figure*}
  \centering
  \subfigure[Energy ratio]{\includegraphics[width=0.475\textwidth]{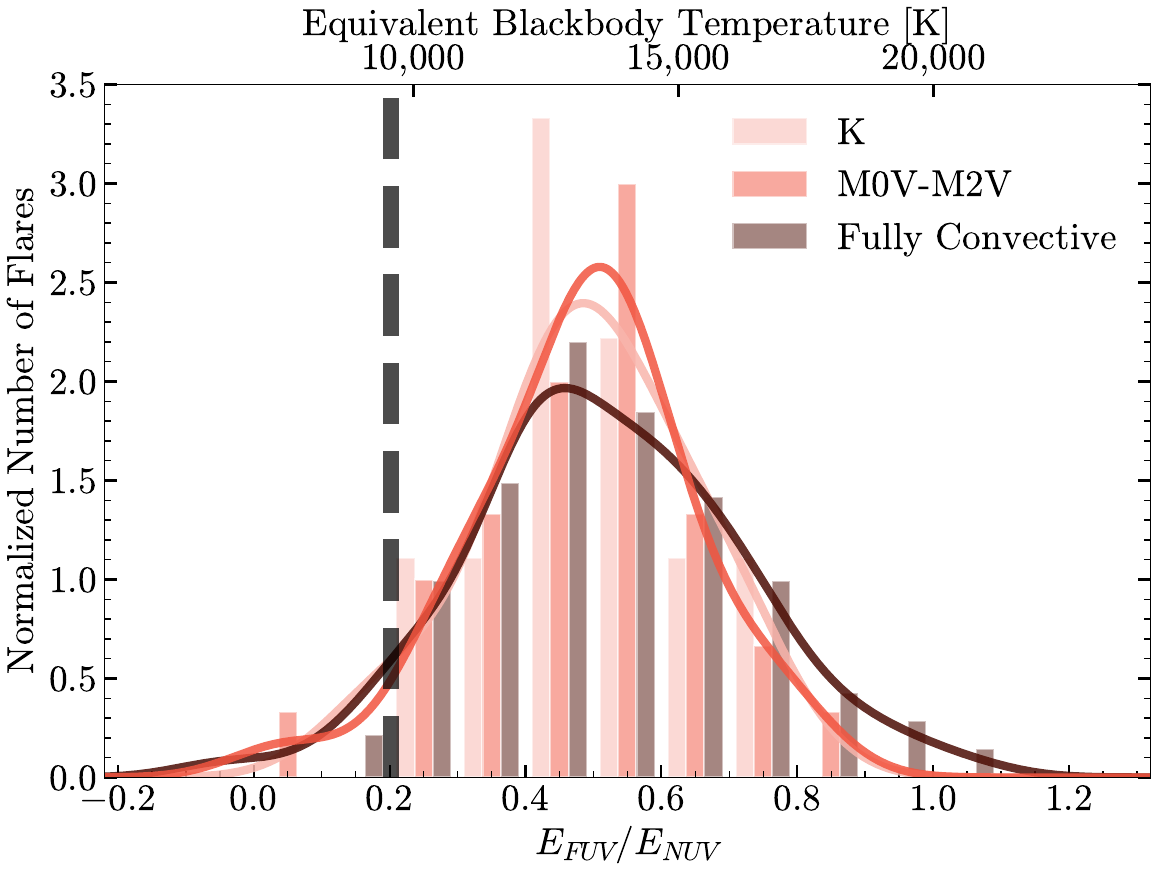}\label{fig:integrated_spt}}
  \hfill
  \subfigure[$\lambda L_{\lambda}$ ratio at peak NUV]{\includegraphics[width=0.475\textwidth]{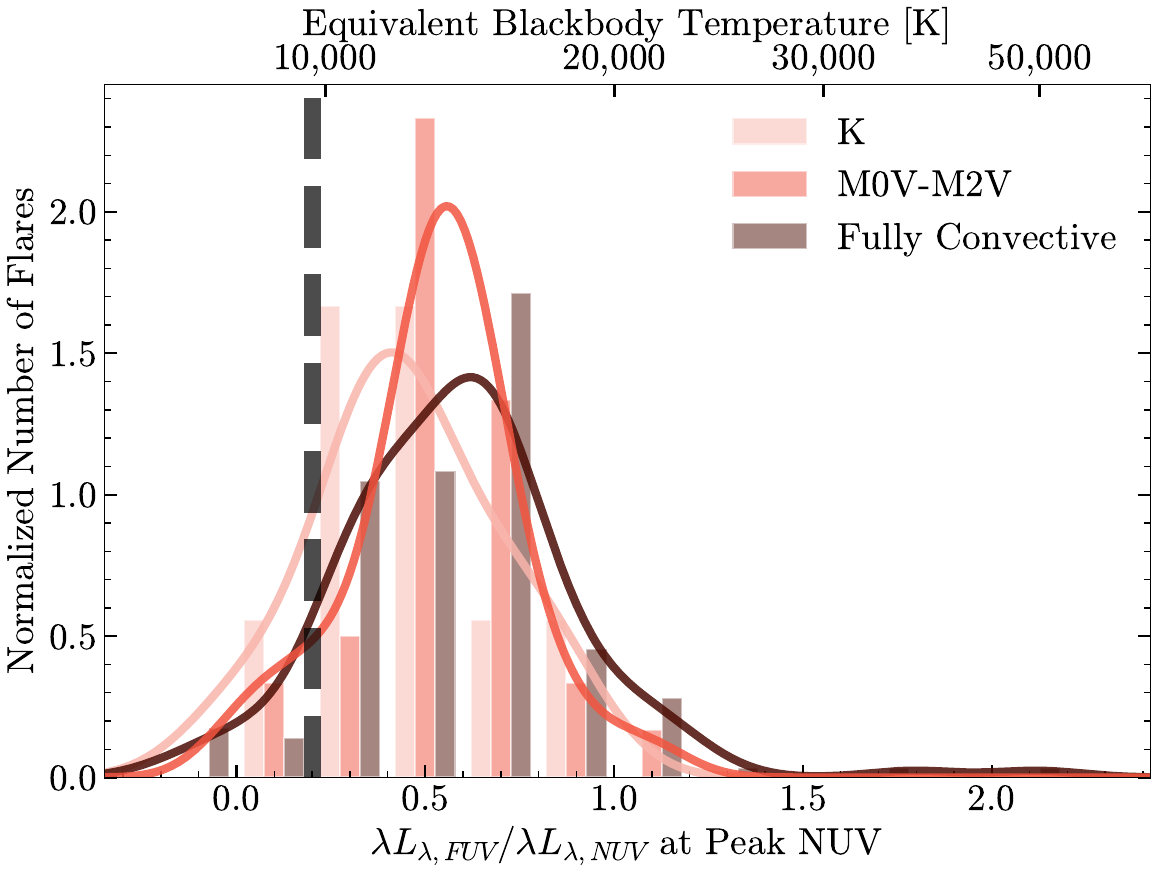}\label{fig:peaknuv_spt}}
  \vspace{1em}
  \subfigure[Bootstrapped energy ratio]{\includegraphics[width=0.475\textwidth]{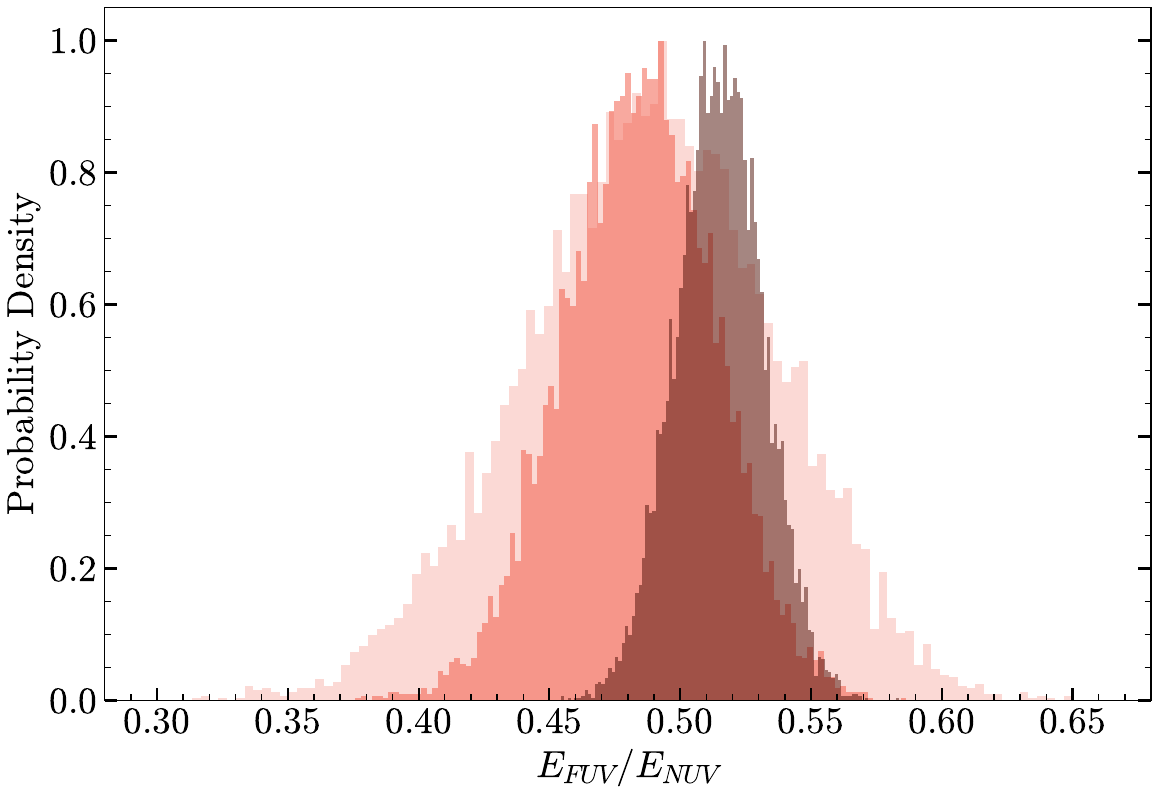}\label{fig:bs_integrated_spt}}
  \hfill
  \subfigure[Bootstrapped $\lambda L_{\lambda}$ ratio]{\includegraphics[width=0.475\textwidth]{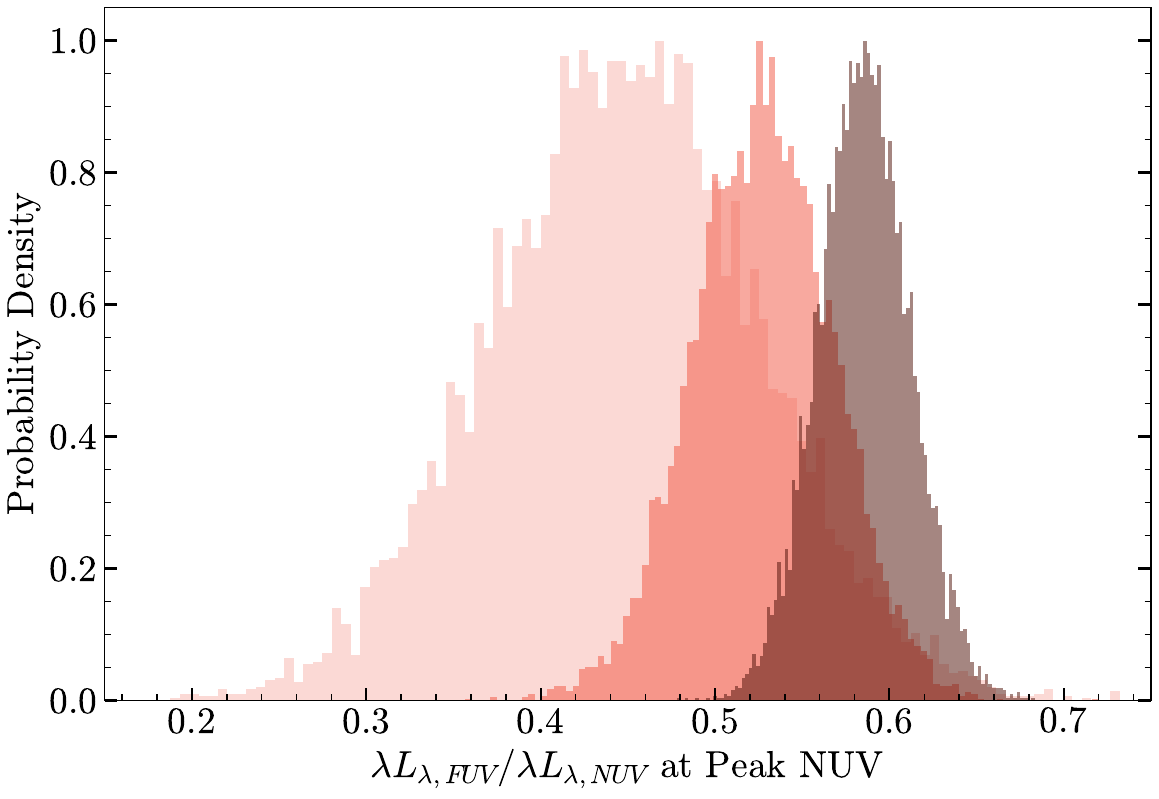}\label{fig:bs_peak_spt}}
  \vspace{1em}
  \subfigure[Energy ratio vs. G-RP color]{\includegraphics[width=0.475\textwidth]{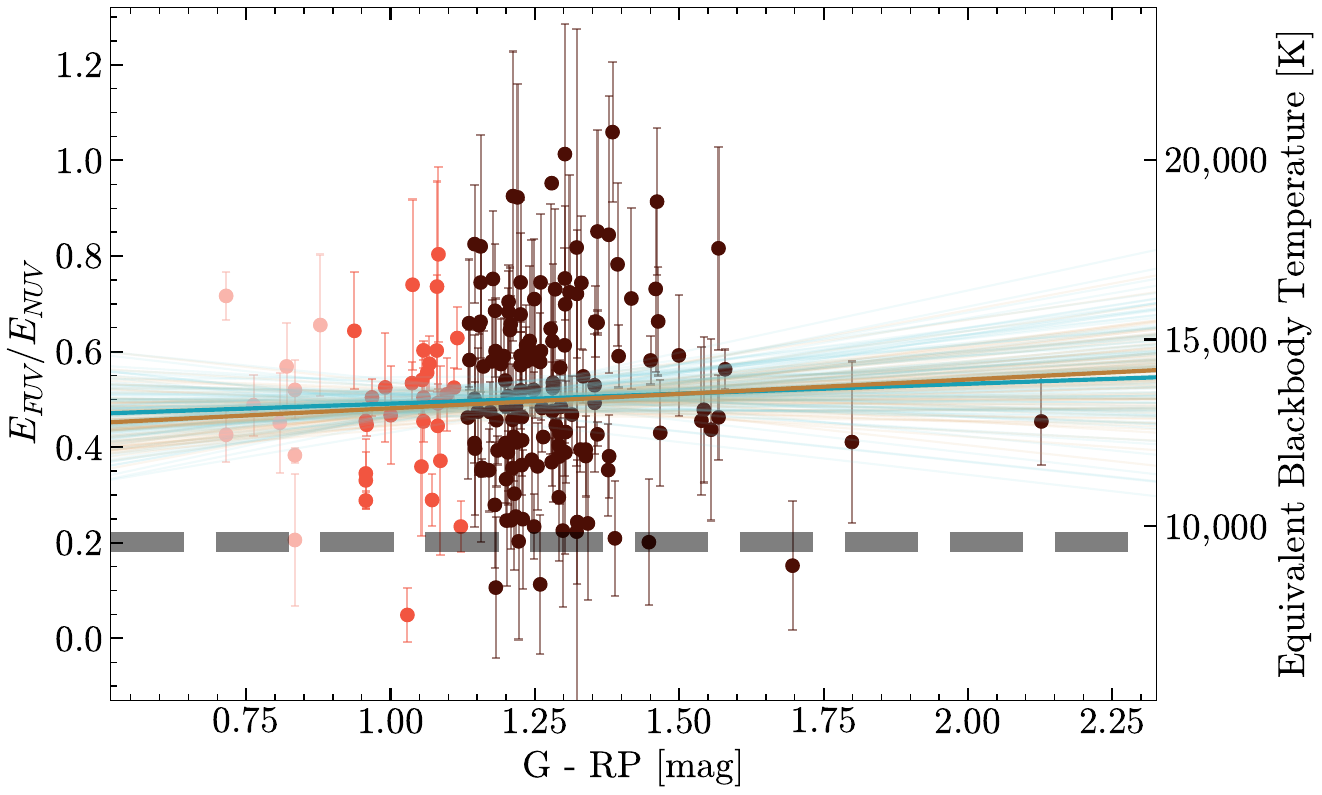}\label{fig:fits_integrated_spt}}
  \hfill
  \subfigure[$\lambda L_{\lambda}$ ratio vs. G-RP color]{\includegraphics[width=0.485\textwidth]{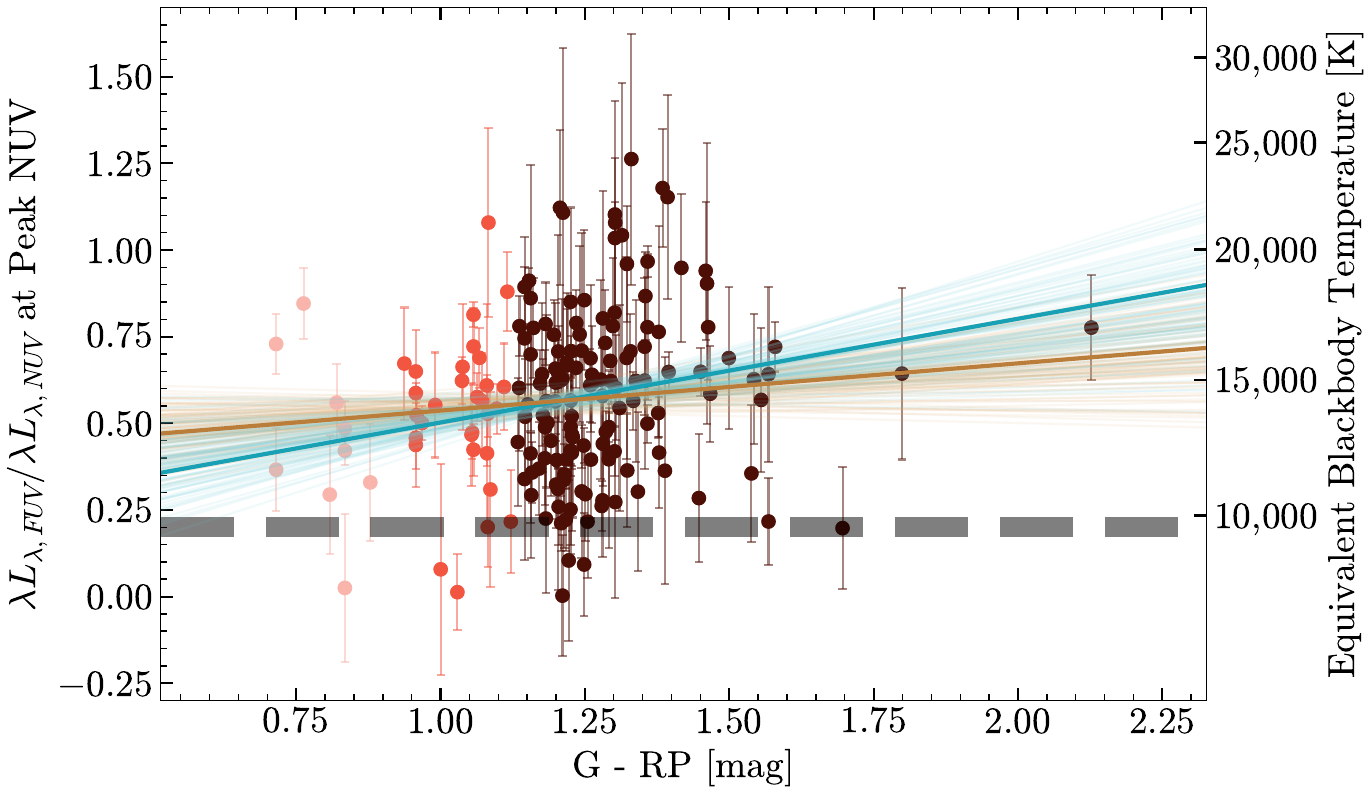}\label{fig:fits_peak_spt}}
  \caption{\small Histograms and scaled kernel density estimates (top), bootstrapped mean values (middle), and scatter plots with lines of best fit (bottom) for FUV/NUV energy ratios (left) and $\lambda L_{\lambda}$ at peak NUV flux (right), with respect to spectral type. The ratios corresponding to $9\,000-10\,000$~K blackbody SEDs are denoted by vertical and horizontal black dashed lines in the top and bottom panels, respectively. Blackbody temperatures corresponding to their respective FUV/NUV ratios are marked on the top axis. X-axes are not consistent between panels. Lines of best fit on the bottom panel are computed using least-squares optimization (orange) and Theil-Sen regression (turquoise). Color darkens with spectral type; ratios for K stars are plotted in light pink, partially convective M stars in red, and fully convective stars in brown.}
  \label{fig:hist_kdes_spt}
\end{figure*}

\begin{figure*}
  \centering
  \subfigure[Energy ratio]{\includegraphics[width=0.475\textwidth]{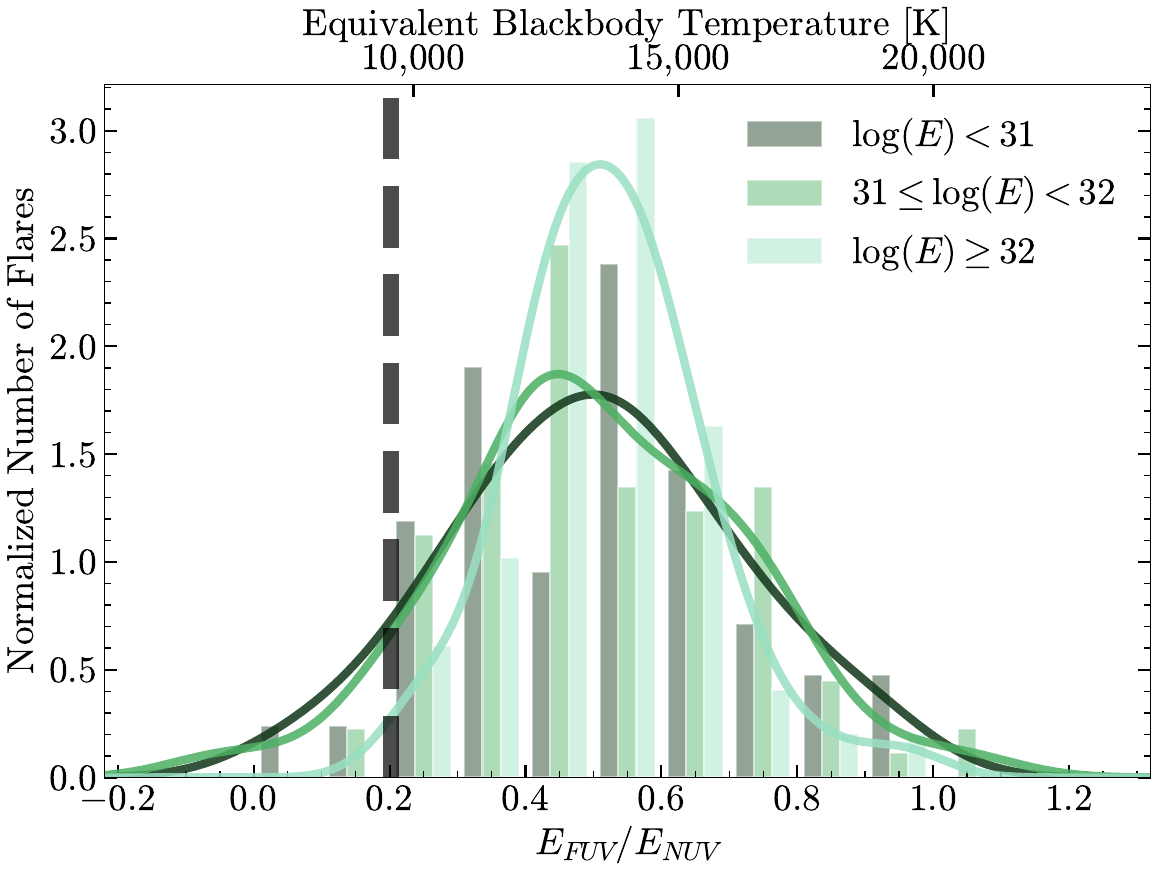}\label{fig:integrated_energy}}
  \hfill
  \subfigure[$\lambda L_{\lambda}$ ratio at peak NUV]{\includegraphics[width=0.48\textwidth]{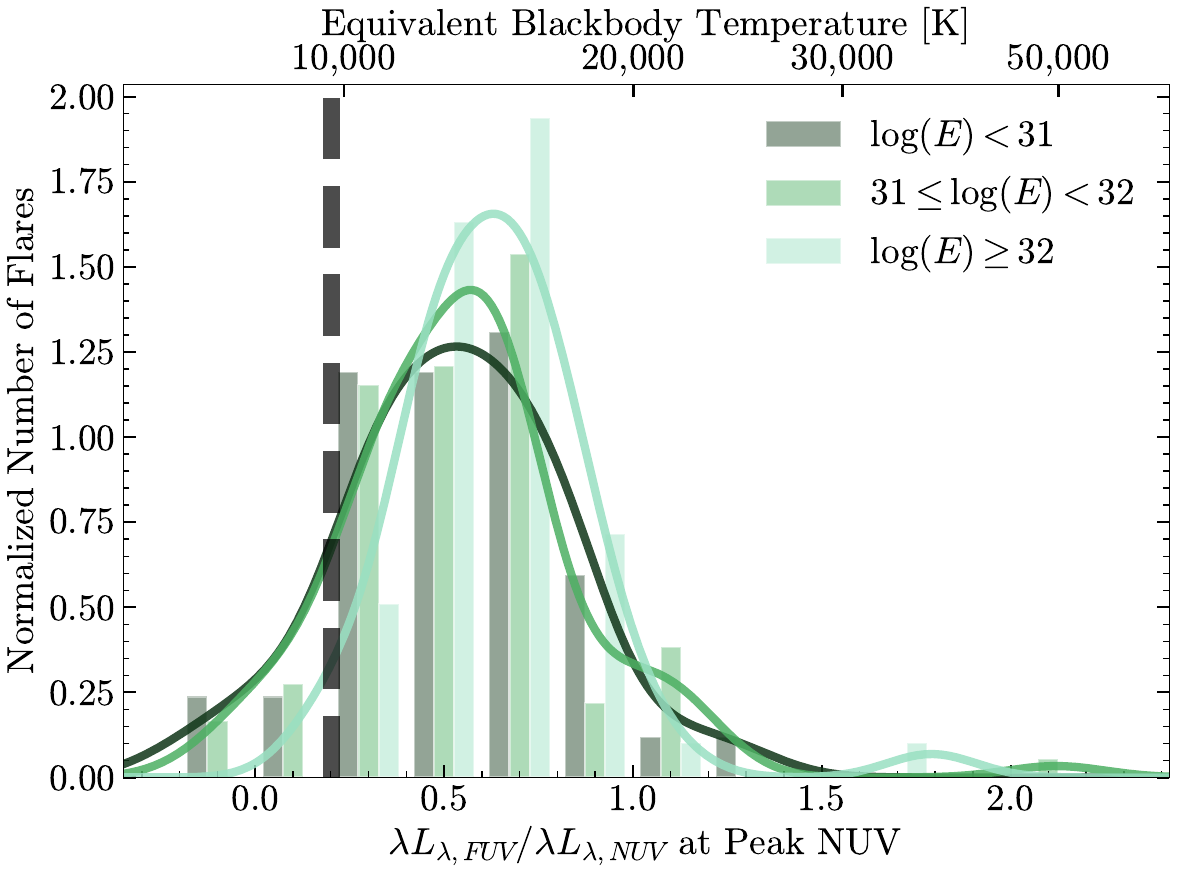}\label{fig:peaknuv_energy}}
  
  \vspace{1em}
  
  \subfigure[Bootstrapped energy ratio]{\includegraphics[width=0.475\textwidth]{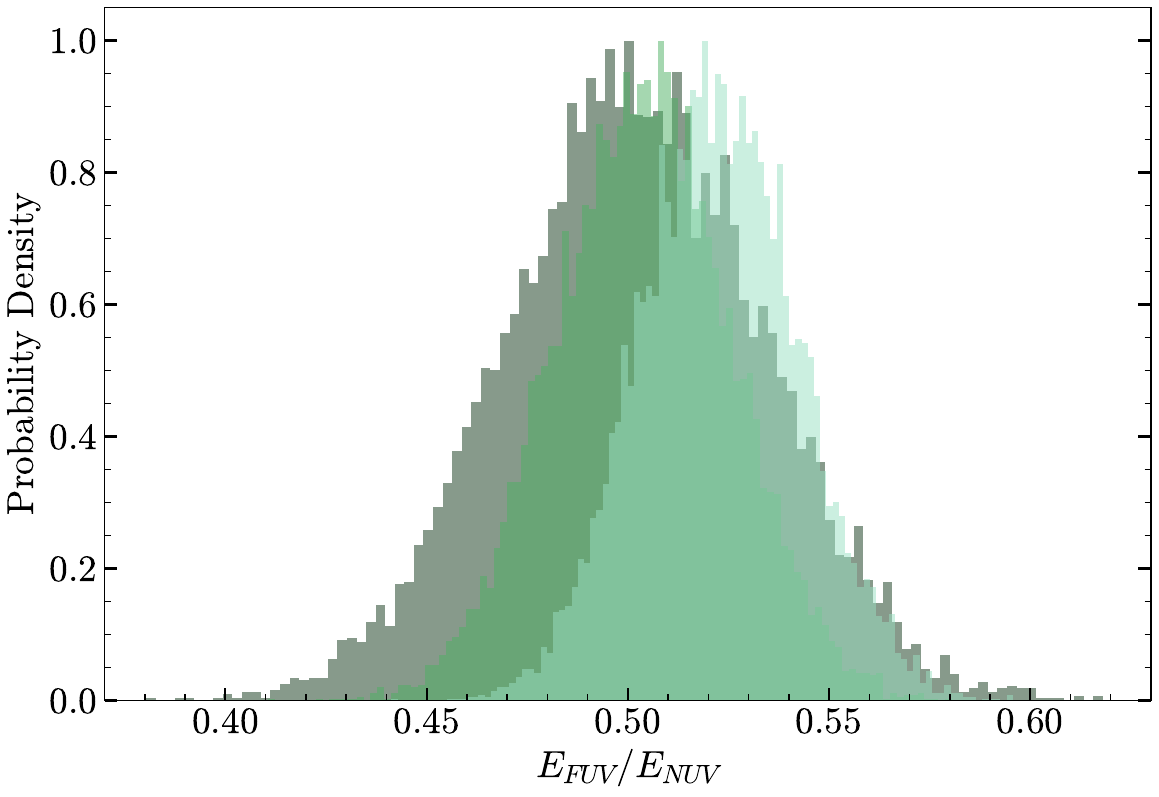}}\label{fig:bs_integrated_energy}
  \hfill
  \subfigure[Bootstrapped $\lambda L_{\lambda}$ ratio]{\includegraphics[width=0.475\textwidth]{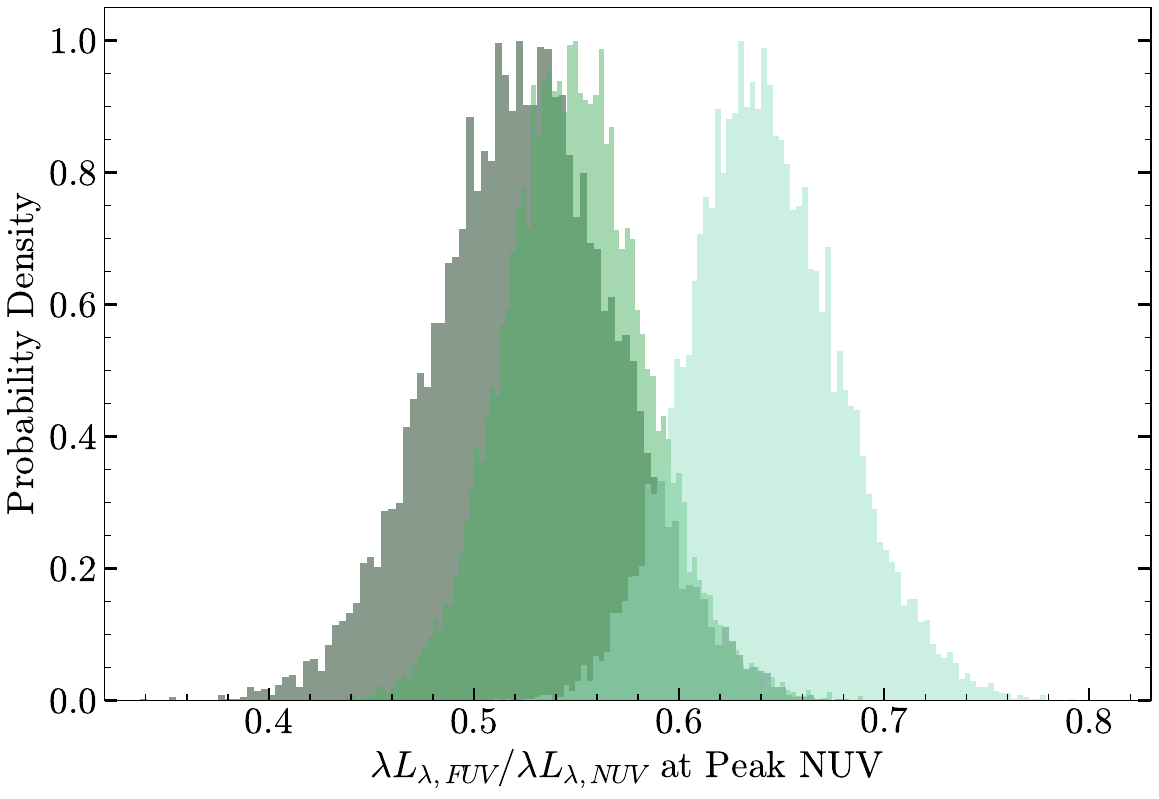}\label{fig:bs_peak_energy}}
  
  \vspace{1em}
  
  \subfigure[Energy ratio vs. log(flare energy)]{\includegraphics[width=0.475\textwidth]{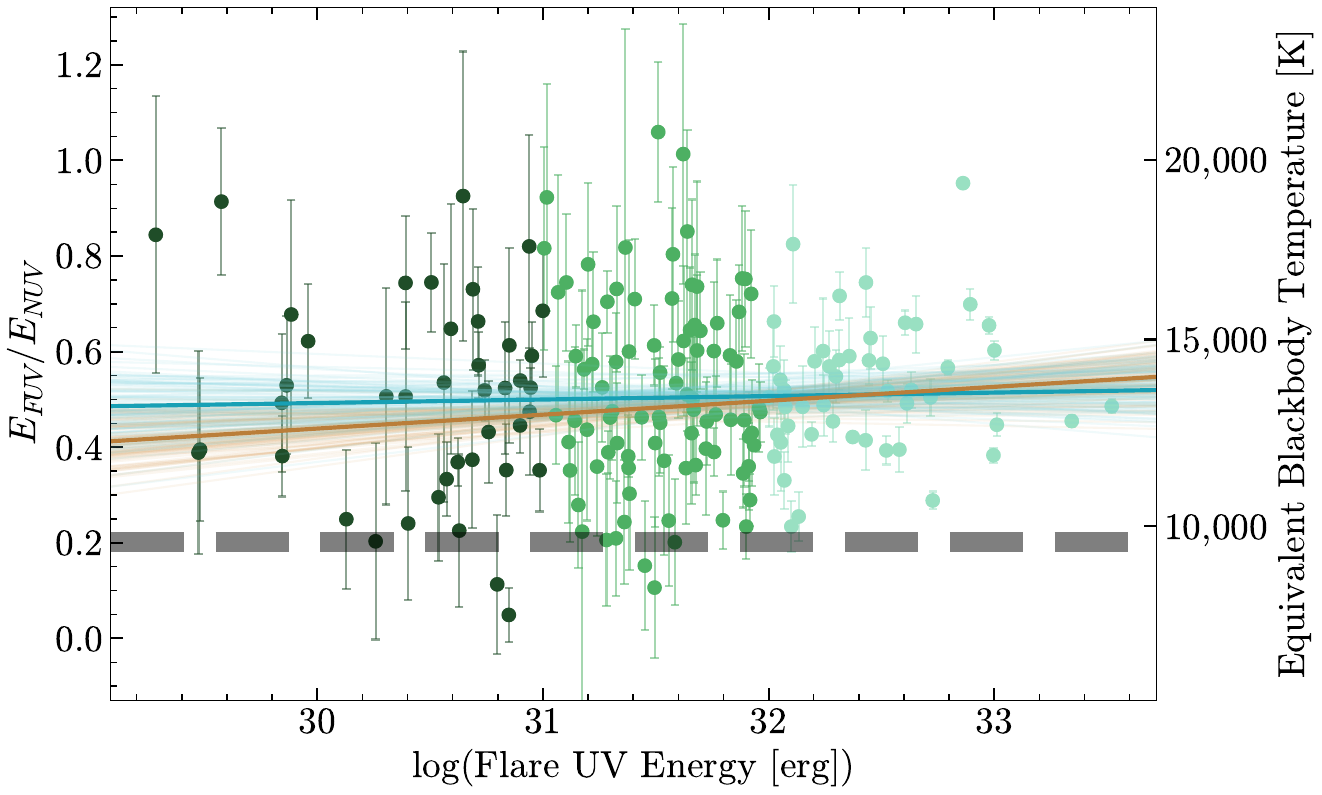}\label{fig:fits_integrated_energy}}
  \hfill
  \subfigure[Energy ratio vs. log(flare energy)]{\includegraphics[width=0.485\textwidth]{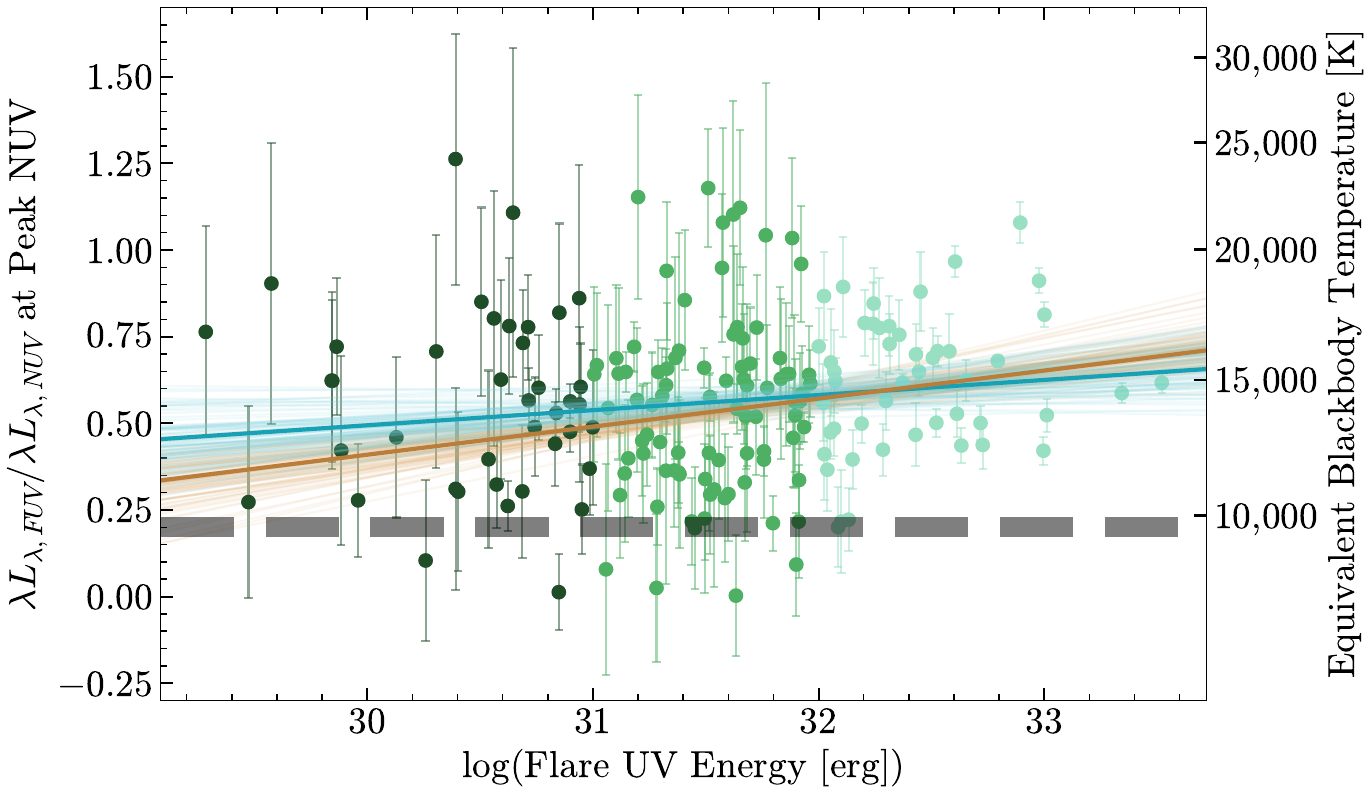}\label{fig:fits_peak_energy}}
  
  \caption{\small Same as Figure \ref{fig:hist_kdes_spt}, but with respect to UV flare energy. Color lightens with an increase in UV flare energy; ratios corresponding to energies below $10^{31}$~erg in dark green, to UV energies between $10^{31}-10^{32}$~erg in medium green, and to energies of $10^{32}$~erg or higher in light green.}
  \label{fig:hist_kdes_energy}
\end{figure*}

On the other hand, FUV/NUV ratios at individual epochs in a flare can significantly exceed the time-integrated values.
To quantify this, we compute the observed ratios of FUV/NUV $\lambda L_{\lambda}$  at peak NUV brightness for each flare, which we refer to as \peakratio.
Figures \ref{fig:peaknuv_spt} and \ref{fig:peaknuv_energy} show the \peakratio{} values with respect to spectral type and flare UV energy.
Note that relying on the NUV time of peak introduces a slight bias against NUV-faint flares.
Thus, true FUV/NUV ratios are likely higher around flare peak.
This behavior is demonstrated in Figure \ref{fig:randflare1}, where the observed FUV flux both exceeds the NUV flux and peaks before the NUV.

We find a median \peakratio = 0.57 ($T_{\rm eff} \approx 14\,300$~K) and a maximum \peakratio = 2.12 ($T_{\rm eff} \approx 49\,600$~K).
Thus, a $9\,000$~K blackbody underestimates the median observed \peakratio{} by a factor of 3.4 and underestimates the maximum ratio by a factor of 12.6.

\subsection{Results by spectral type}\label{3.3}
Stellar magnetic fields, which are the energy sources for stellar flares, originate from dynamo-related convective motions and differential rotation, \citep[e.g.,][]{petterson89, balona15, doorsselaere17} and so the depth of the convection zone may correlate with flare properties. To test this idea, we derive spectral types for the flare stars in our sample using Gaia G-RP color \citep{gaiaedr3} and the spectral type-color sequence from \citet{pecaut13spt}, estimating a fully convective boundary of M3V \citep{jao18}. We use this system because we do not have rotation periods for every star in the sample to compute a Rossby number.
We divide our sample into three spectral classification bins: K stars, partially convective M dwarfs (M0V-M2V), and fully convective stars (M3$+$).
Figures \ref{fig:integrated_spt} and \ref{fig:peaknuv_spt} illustrate the distributions of time-integrated energy ratios \energyratio{} and \peakratio{} by spectral bin.  Table \ref{tab:kdes} displays the medians and standard deviations for each kernel density estimate (KDE).

There appears to be a trend with an increasing median \peakratio{} with later spectral types and thus deepening convective envelopes.
The distribution of \peakratio{} for K stars has a median of 0.443 ($12\,700$~K); M0V-M2V stars a median \peakratio{} of 0.541 ($14\,000$~K); and fully convective stars a median \peakratio{} of 0.577 ($14\,400$~K). 
The \energyratio{} distributions peak at similar values but skew towards higher ratios, a similar trend to \peakratio.
Furthermore, fully convective stars exhibit the largest deviations from the assumed $9\,000$~K blackbody. 

We explore the significance of these findings in three ways:

\noindent 1) As a first attempt, we pairwise compared the spectral type categories using Kolmogorov-Smirnov (K-S) and Anderson-Darling tests.  Neither of these tests were able to differentiate either the \energyratio{} or \peakratio{} distributions at 95\% confidence. 

\noindent 2) As a next attempt, we bootstrapped the mean ratios for each spectral type category. We resampled each category with replacement $10\,000$ times and calculated the mean of the distribution.  We compute \energyratio{} means and quantile intervals for K, early M, and fully convective stars of $0.49\pm0.05$,  $0.49\pm0.03$, and $0.52\pm0.02$. We compute \peakratio{} means and quantile intervals for K, early M, and fully convective stars of $0.45\pm0.08$,  $0.53\pm0.04$,  and $0.59^{+0.02}_{-0.03}$, respectively.  The appears to be a statistically significant correlation between \peakratio{} and spectral type.

\noindent 3) Finally, we explore a potential linear relationship between the FUV/NUV ratios and $G$-$RP$ color as a proxy for spectral type (shown in Figure~\ref{fig:fits_integrated_spt} and \ref{fig:fits_peak_spt}).  As a first step, we cut obvious outliers, five with negative \peakratio{} and two with large but very uncertain \peakratio.
The Pearson correlation coefficient is 0.18 for \peakratio{} vs. $G$-$RP$ and 0.06 for \energyratio{} vs. $G$-$RP$.  These indicate a tentative weak correlation but the significance is difficult to determine because the Pearson correlation coefficient does not consider uncertainties.  
Instead, we will perform a linear fit and bootstrap the sample to determine if there is a statistically significant positive slope.   First, to reduce the covariance between the slope and intercept we move the intercept to the median color of the sample as given by:
\begin{equation}
    \mathcal{R} = m \times ({\rm log_{10} (E_{\text{UV}}/{\rm erg}}) - 1.2) + b,
   \label{eq:line}
\end{equation}
\noindent  for slope $m$, intercept $b$, and ratio $\mathcal{R}$ (either \energyratio{} or \peakratio{}). Second, to account both for underestimated errors and for the intrinsic scatter in the flare population, we iteratively fit a best-fit line adding increasing uncertainty in quadrature with the \energyratio{} or \peakratio{} uncertainties until a reduced $\chi^2=1$  was found for each fit. We report the best-fit values in Table~\ref{tab:fits} and show the best-fits in Figures~\ref{fig:fits_integrated_spt} and \ref{fig:fits_peak_spt}. Next we determine the uncertainty on the linear fit by bootstrapping the sample with replacement $10\,000$ times.  We show 100 randomly selected bootstrapped fits in Figures~\ref{fig:fits_integrated_spt} and \ref{fig:fits_peak_spt} to indicate the fit uncertainty. In Table~\ref{tab:fits} we report the 16$^{th}$ and 84$^{th}$ percentile along with the percentage of the resamples where the slope of the best fit is $\leq 0$.
Finally, to verify the least-square fits are not being driven by outliers, we redo the linear fits using Theil-Sen Regression, bootstrap the fits, and also report those fits in Table~\ref{tab:fits}. We then take the fraction of bootstrap trials with slopes $\leq$ 0 as our p-value for the detection of a positive relation.  When reporting the significance, we take the more conservative value between the least-square and Theil-Sen fits.
We find that the \peakratio{} vs. color slope is consistent with being positively correlated at nearly 2 sigma and the \energyratio{} vs. color slope is consistent with being positively correlated at nearly 1.5 sigma.

Thus we conclude there is a tentative trend with increasing average \peakratio{}  and \energyratio{} with later spectral types, especially for \peakratio{} and for fully convective stars.  However, we note that the uncertainties of the \peakratio{}  and \energyratio{} ratios are large relative to the difference in the means between spectral type categories.

\subsection{Results by flare UV energy}
We also investigate for trends between the total flare UV energy and FUV/NUV ratio. First, we estimate the total UV flux at each epoch by fitting a linear SED between ($\lambda_{\text{eff}}^{\text{FUV}}, f_{\text{FUV}}$)  and ($\lambda_{\text{eff}}^{\text{NUV}}, f_{\text{NUV}}$) for an observed FUV flux $f_{\text{FUV}}$ and corresponding NUV flux $f_{\text{NUV}}$ and effective wavelengths $\lambda_{\text{eff}}^{\text{FUV}}=1528$~\AA{}  and $\lambda_{\text{eff}}^{\text{NUV}}=2271$~\AA{} \citep{morrissey05}. We integrate the linear SED over the wavelength range covered by 10\% of peak filter response for the \emph{GALEX} filters, $1343\-- 2831$~\AA{}, as shown in Figure \ref{fig:filter_resp}.  We then use the distance from Gaia as presented in Table~\ref{tab:stars} to compute total flare UV energies ($E_{\text{UV}}$).
We find that our flare sample has $E_{\text{UV}} = 2\times10^{29} \-- 3\times10^{33}$ erg, with a median UV energy of $3.8 \times 10^{31}$ erg.

We divide the flare sample into 3 bins of $E_{\text{UV}}$, flares with $E_{\text{UV}}<10^{31}$~erg, $10^{31}<E_{\text{UV}}< 10^{32}$~erg, and $E_{\text{UV}}>10^{32}$~erg. 
Figures \ref{fig:integrated_energy} and \ref{fig:peaknuv_energy} display  \energyratio{} and \peakratio{} as a function of $E_{\text{UV}}$.  
The corresponding values of the kernel density estimates by UV energy are noted in Table \ref{tab:kdes}. 
The distribution of \peakratio{} for flare UV energies below $10^{31}$~erg has a median of 0.530 ($13\,800$~K); energies between $10^{31}-10^{32}$~erg a median of 0.535 ($13\,900$~K); and energies $10^{32}$~erg and above a median of 0.628 ($15\,000$~K).

 We then perform similar statistical tests as in Section~\ref{3.3}.  

\noindent 1) K-S and Anderson Darling tests again can mostly not distinguish the \energyratio{} or \peakratio{} distributions for the $E_{\text{UV}}$ groups.  However, the K-S test between the $10^{31}<E_{\text{UV}}< 10^{32}$~erg and $E_{\text{UV}}>10^{32}$~erg \peakratio{} samples has a marginal, p-value of 0.03, detection that they are drawn from different distributions. 

\noindent 2) 
We compute \energyratio{} means and quantile intervals for flare energies of $E_{\text{UV}}<10^{31}$~erg, $10^{31}<E_{\text{UV}}< 10^{32}$~erg, and $E_{\text{UV}}>10^{32}$~erg of $0.50\pm0.03$, $0.51\pm0.02$, and $0.52\pm0.02$.
We compute \peakratio{} means and quantile intervals of $0.53\pm0.04$,  $0.55\pm0.03$,  and $0.64\pm0.04$, respectively.
There appears to be a statistically significant trend where the most energetic flares have a higher \peakratio{} but it does not appear such a relation exists for \energyratio{}.

\noindent 3)
We follow the same procedure as in Section~3.3 investigating if there is a linear relationship between FUV/NUV ratios and $E_{\text{UV}}$.
First, the Pearson correlation coefficient is 0.11 for \peakratio{} vs. $E_{\text{UV}}$ and 0.02 for \energyratio{} vs. $E_{\text{UV}}$.  These indicate a tentative weak correlation but we bootstrap least-squares and Theil-Sen linear fits to test their statistical significance. 
Again to reduce covariance between the fit parameters we choose a pivot near the median $E_{\text{UV}}$ as shown by:
\begin{equation}
\mathcal{R} = m \times ({\rm log_{10} (E_{\text{UV}}/{\rm erg}}) - 31.6) + b,
   \label{eq:Eline}
\end{equation}
\noindent for slope $m$, intercept $b$, and ratio $\mathcal{R}$. The best-fits are shown in Figures~\ref{fig:fits_integrated_energy} and \ref{fig:fits_peak_energy}. In Table~\ref{tab:fits} we report the fit statistics. 
We find that the \peakratio{} vs. $E_{\text{UV}}$ slope is consistent with being positively correlated at $\gtrsim 2$ sigma.  However, the outlier-resistant Theil-Sen fit is consistent with no relationship between \energyratio{} and $E_{\text{UV}}$. 

Thus we conclude there is a trend with increasing  \peakratio{} for more UV luminous flares, especially for the largest ($E_{\text{UV}}>10^{32}$~erg) flares. However, we do not find evidence for a similar trend with \energyratio{}. 

\begin{table*}[h]
    \centering
    \caption{Kernel Density Estimates}
    \begin{tabular}{ p{3cm} | p{2cm} p{2cm} p{2cm} p{2.5cm} p{2cm} p{2cm}}
    \hhline{=======}
&  \multirow{2}{3em}{\energyratio{} median} & \multirow{2}{3em}{16th percentile} & \multirow{2}{3em}{84th percentile} & \multirow{2}{7.5em}{\peakratio{} median}& \multirow{2}{3em}{16th percentile} & \multirow{2}{3em}{84th percentile} \\ 
\hline
$\log(E_{UV}<31)$ & 0.501 & 0.277 & 0.734 & 0.530 & 0.219 & 0.833 \\ 
$31 \leq \log(E_{UV})<32$ & 0.497 & 0.289 & 0.739 & 0.535 & 0.242 & 0.829 \\ 
$\log(E_{UV})\geq 32$ & 0.518 & 0.382 & 0.662 & 0.628 & 0.395 & 0.865\\ 
K & 0.494 & 0.320 & 0.666 & 0.443 & 0.178 & 0.739 \\ 
Early M & 0.493 & 0.318 & 0.652 & 0.541 & 0.301 & 0.743\\ 
Fully Convective & 0.509 & 0.313 & 0.725 & 0.577 & 0.284 & 0.863\\ 
\hline 
    \end{tabular}
    \label{tab:kdes}
\end{table*}

\begin{table*}[h]
    \centering
    \caption{Fits to FUV/NUV trends with respect to G-RP color and flare UV energy. 
    }
    \begin{tabular}{ p{3cm} | p{1.1cm} p{1.1cm} p{2cm} p{2cm} p{2cm} p{2cm} p{2cm}}
    \hhline{========}
&  \multirow{2}{4em}{Best-fit slope} & \multirow{2}{4em}{Best-fit intercept} & \multirow{2}{7em}{16th percentile slope} & \multirow{2}{7em}{84th percentile slope} & \multirow{2}{7em}{16th percentile intercept} & \multirow{2}{7em}{84th percentile intercept} & \multirow{2}{7em}{\% slope $\leq$ 0}  \\ 
&  \multirow{2}{4em}{Best-fit $m$} & \multirow{2}{4em}{Best-fit ~ $b$} & \multirow{2}{7em}{16th percentile $m$} & \multirow{2}{7em}{84th percentile $m$} & \multirow{2}{7em}{16th percentile $b$} & \multirow{2}{7em}{84th percentile $b$} & \multirow{2}{7em}{\% $m$ $\leq$ 0}  \\ 
\hline
Least-squares fit \\
\hline
\peakratio{} vs. G-RP & 0.14 & 0.56 & 0.05 & 0.23 &  0.55 & 0.58 & \hphantom{1}6.80 \\ 
\peakratio{} vs. UV energy & 0.08 & 0.54 & 0.05 & 0.10  & 0.52 & 0.56 & \hphantom{1}0.04 \\ 
\energyratio{} vs. G-RP & 0.06 & 0.49 & 0.00 & 0.13  & 0.48 & 0.51 & 16.25 \\ 
\energyratio{} vs. UV energy & 0.03 & 0.49 & 0.01 & 0.05 &  0.47 & 0.50 & \hphantom{1}4.65 \\ 
\hline
Theil-Sen fit \\
\hline 
\peakratio{} vs. G-RP & 0.24 & 0.56 & 0.13 & 0.30 &  0.55 & 0.58 & \hphantom{1}0.30 \\ 
\peakratio{} vs. UV energy & 0.04 & 0.56 & 0.02 & 0.07  & 0.54 & 0.58 & \hphantom{1}3.80 \\ 
\energyratio{} vs. G-RP & 0.09 & 0.50 & 0.00 & 0.16 & 0.49 & 0.52 & 18.95 \\ 
\energyratio{} vs. UV energy & 0.00 & 0.50 & 0.00 & 0.03 & 0.48  & 0.52 & 31.15 \\ 
\hline 
    \end{tabular}
    \label{tab:fits}
\end{table*}

\section{Potential Impacts on Habitability}\label{s4:discussion}

We have demonstrated that uniformly selected field stars exhibit flares that are FUV luminous and not well-represented by a constant $9\,000 - 10\,000$~K blackbody. This finding holds for observed fluxes, integrated energy and peak luminosity. In this section, we discuss some implications of this finding.

Ultraviolet-C radiation (UVC, $\sim$2000-2800 \AA) is expected to drive prebiotic chemistry  \citep{todd18, rimmer18} and ozone depletion  \citep{segura10, tilley19}. If blackbody temperatures during flares indeed exceed  $9\,000$~K, then flares deliver higher levels of NUV flare radiation than assumed when characterizing the impact of flaring on abiogenesis and atmospheric ozone depletion.

Optical studies of flares and their potential impact on exoplanet habitability typically assume a constant $9\,000$~K blackbody for flares, which impacts their calculation of flare energy 
\citep{schmidt19, rodriguez18, gunther20, zeldes21, bogner21}. 
\citet{zeldes21} selected a sample of five superflaring stars and found that none fell into either zone.

In a sample of 1228 flaring stars from TESS, \citet{gunther20} found eight percent of sources to display sufficient flaring rates and energies to fall into the ozone depletion zone, and only one percent to fall in the abiogenesis zone. 
If instead these flares are well-represented by $13\,500$ K blackbody SED, the number of stars with sufficient flaring activity to fall in the ozone depletion zone would increase by 60\%. In particular, the number of early M stars (M0-M4) in the ozone depletion zone would increase by 270\%. 

Characterizing the impact of higher flare temperatures on prebiotic chemistry is less straightforward. \citet{rimmer18} assumes an Earth-like atmosphere which will absorb wavelengths shorter than 2100~\AA{}, and focuses on NUV radiation ($2000-2800$~\AA) as a driver of prebiotic chemistry. \citet{rimmer18} estimates the total number of photons between $2000-2800$~\AA{} reaching a planet's surface by assuming the AD Leo flare spectrum in their model is flat at wavelengths shorter than the U band. We found this is likely not the case in the FUV ($1350-1750$~\AA) and NUV ($1750-2750$~\AA). This calls into question their conclusion that the number of NUV photons reaching a planet's surface can be inferred from the U-band flare energy. 

In future work (Berger et al. 2024, in prep) we will directly compute population-averaged NUV flare rate distributions, directly probing the wavelength ranges thought to be important for the formation of complex molecules. Naively scaling a blackbody in TESS from 9000 K to $13\,500$~K will result in an increase in NUV photons by a factor 3.86.

UV radiation from stars impacts the atmospheric composition of orbiting planets.
Large ratios of FUV/NUV radiation as seen on M dwarfs may produce sufficient abiotic atmospheric oxygen to produce a false biosignature \citep{tian14, harman15}.  
These studies characterize UV radiation in steady state and do not account for flaring events.
Additional modeling work should be undertaken to investigate if the stronger FUV flux during flares that we observe in this work can	significantly alter atmospheric oxygen abundances and thus affecting interpretation of exo-atmospheric signatures. A few FUV spectra of flares exist \citep{hawley91, france16, loyd18a, froning19} but the short timescales of flares make obtaining these observations difficult.

We identified no observations with FUV/NUV flux ratios above the 4.33 \energyratio{} asymptote in our color-temperature curve (Figure \ref{fig:color-temp}) above which the overall FUV emission could not be attributed solely to blackbody  emission. 
Thus for the stars in our sample, we cannot determine the extent to which line emission contributes to the total FUV emission we observe. 
To characterize the respective contributions of line and continuum emission to overall FUV flare emission will require
obtaining FUV spectra of a population of stars in quiescence and flare.  Spectral observations of a large solar flare suggest that line emission can indeed dominate the overall UV flare emission \citep{simoes19}.
Our results highlight the importance of simultaneous observations of flares across the electromagnetic spectrum in order to fully capture the emission processes involved \citep[e.g.,][]{paudel21}.

Looking forward, two missions planned for launch in 2026 may advance our understanding of stellar flares in the ultraviolet.
The Ultraviolet Transient Astronomy Satellite \citep[ULTRASAT;][]{ultrasat23}  plans to observe $>10^5$ flaring and variable stars in the NUV ($2300-2900$~\AA), sensitive to $<23.5$~AB mag. 
ULTRASAT will provide insight into the levels of NUV radiation from flares that may drive prebiotic chemistry or atmospheric ozone depletion. 
However, the single-filter design of ULTRASAT will not allow any temperature estimates.
Additionally, the Monitoring Activity of Nearby sTars with uv Imaging and Spectroscopy \citep[MANTIS;][]{mantisprop22}   mission will work alongside JWST and may be a better probe of flare energetics. MANTIS will
observe cool stars in the extreme-ultraviolet ($100-900$~\AA), far-ultraviolet ($900-2000$~\AA), near-ultraviolet ($2000-3200$~\AA), and visible ($3200-10\,000$~\AA) wavelengths, providing a detailed estimate of the color temperature.
MANTIS will provide the first glimpse at many stars in the EUV, and may shed light on the proportion of flare radiation originating from continuum versus UV line emission.

\section{Conclusions}\label{s5:conclusions}

In this work, we have identified 182 flares on 158 stars with simultaneous observations in the NUV and FUV from the \galex \, space telescope.  
We select our targets from the Gaia Catalogue of Nearby Stars which is complete down to spectral type M8 within our survey volume of 100 pc.  Thus, this study suffers from minimal selection basis. 
We have computed color temperatures at each epoch of the flares in our sample using a derived relation between FUV/NUV flux and blackbody temperature. 
Our main results are as follows.

\begin{enumerate}
    \item A constant $9\,000$~K blackbody underpredicts the FUV emission for 98\% of flares in our sample. 
    
    \item The FUV/NUV ratio at peak appears to positively correlate ($\sim2 \sigma$ significance) both with total UV flare energy and with increasing $G - RP$ color of the host star.
    
\end{enumerate}

Future work (Berger in prep.) will use this \galex{} flare sample and an injection and recovery analysis to compute the UV flare rate distribution and produce a UV flare model.

% --------------- ACKNOWLEDGEMENTS ---------------------------------------
\section*{Acknowledgements}

VLB acknowledges support from the Research Experience for Undergraduates program at the Institute for Astronomy, University of Hawaii-Manoa funded through NSF grant \#2050710. VLB thanks the Institute of Astronomy for its hospitality over the course of this project. VLB is also supported by a Churchill Scholarship.
JTH was supported by NASA grant 80NSSC21K0136.
BJS is supported by NSF grants AST-1907570, AST-1920392 and AST-1911074. 
JvS acknowledges support from NASA grant 80NSSC22K0293.
D.H. acknowledges support from the Alfred P. Sloan Foundation and the Australian Research Council (FT200100871).
We thank Dr. Michael Liu and Dr. Allison Youngblood for useful discussions.

Software: \texttt{matplotlib} \citep{hunter07}, \texttt{numpy} \citep{harris20}, \texttt{pandas} \citep{mckinney10}, \texttt{scipy} \citep{scipy}, \texttt{astropy} \citep{astropy:2013, astropy:2018, astropy:2022}.

% --------------- DATA AVAILABILITY ---------------------------------------
\section*{Data Availability}
This manuscript only uses publicly available data from \emph{GALEX} and \emph{Gaia}.
All derived data products will be made publicly available upon publication.

% --------------- BIBLIOGRAPHY ---------------------------------------
\bibliography{bibliography.bib}
\bibliographystyle{mnras}

\end{document}